\documentclass[final,compsoc,twocolumn]{IEEEtran}
\usepackage[pdftex]{graphicx}
\usepackage{epstopdf}
\usepackage[cmex10]{amsmath}
\usepackage{cases,amssymb}
\usepackage{subeqnarray}
\usepackage[linesnumbered,ruled,vlined]{algorithm2e}
\usepackage{cite,setspace,bm,enumerate}
\usepackage{amsmath,color}
\usepackage{amsfonts,caption}
\usepackage{mathrsfs}
\usepackage{algpseudocode}
\usepackage{chngpage}
\usepackage{CJKutf8}
\usepackage{algorithmicx}

\usepackage{url}

\usepackage{soul}
\usepackage{caption}
\usepackage{graphicx}
\usepackage{float} 
\usepackage{subfig} 
\usepackage{color}
\usepackage{booktabs}
\usepackage{multirow}

\newtheorem{definition}{\textbf{Definition}}

\newtheorem{lemma}{\textbf{Lemma}}
\newtheorem{theorem}{\textbf{Theorem}}

\begin{document}
\title{Federated Digital Twin Construction via\\Distributed Sensing: A Game-Theoretic Online Optimization with Overlapping Coalitions}

\author{Ruoyang~Chen,
	Changyan~Yi,~\IEEEmembership{Senior Member,~IEEE},
	Fuhui~Zhou,~\IEEEmembership{Senior Member,~IEEE},
	Jiawen~Kang,~\IEEEmembership{Senior Member,~IEEE},
	Yuan~Wu,~\IEEEmembership{Senior Member,~IEEE},
	Dusit~Niyato,~\IEEEmembership{Fellow,~IEEE}
	\IEEEcompsocitemizethanks{
		\IEEEcompsocthanksitem R. Chen and C. Yi are with the College of Computer Science and Technology, Nanjing University of Aeronautics and Astronautics, Nanjing, 211106, China. (e-mail: \{ruoyangchen, changyan.yi\}@nuaa.edu.cn).
		\IEEEcompsocthanksitem F. Zhou is with the College of Electronic and Information Engineering, Nanjing University of Astronautics and Aeronautics, Nanjing 210016, China (e-mail: zhoufuhui@ieee.org).
		\IEEEcompsocthanksitem J. Kang is with the School of Automation, Guangdong University of Technology, Guangzhou 510006, China (e-mail: kavinkang@gdut.edu.cn).
		\IEEEcompsocthanksitem Y. Wu is with the State Key Laboratory of Internet of Things for Smart City, University of Macau, Macau, China, and also with the Department of Computer and Information Science, University of Macau, Macau, China (e-mail: yuanwu@um.edu.mo).
		\IEEEcompsocthanksitem D. Niyato is with the College of Computing and Data Science, Nanyang Technological University, Singapore 639798 (e-mail: dniyato@ntu.edu.sg).
	}
}

\IEEEtitleabstractindextext{%
\begin{abstract}
In this paper, we propose a novel federated framework for constructing the digital twin (DT) model, referring to a living and self-evolving visualization model empowered by artificial intelligence, enabled by distributed sensing under edge-cloud collaboration. In this framework, the DT model to be built at the cloud is regarded as a global one being split into and integrating from multiple functional components, i.e., partial-DTs, created at various edge servers (ESs) using feature data collected by \textcolor{black}{associated sensors}. Considering time-varying DT evolutions and heterogeneities among partial-DTs, we formulate an online problem that jointly and dynamically optimizes partial-DT assignments from the cloud to ESs, ES-sensor associations for partial-DT creation, and as well as computation and communication resource allocations for global-DT integration. The problem aims to maximize the constructed DT’s model quality while minimizing all induced costs, including energy consumption and configuration costs, in long runs. To this end, we first transform the original problem into an equivalent hierarchical game with an upper-layer two-sided matching game and a lower-layer overlapping coalition formation game. After analyzing these games in detail, we apply the Gale-Shapley algorithm and particularly develop a switch rules-based overlapping coalition formation algorithm to obtain short-term equilibria of upper-layer and lower-layer subgames, respectively. Then, we design a deep reinforcement learning-based solution, called DMO, to extend the result into a long-term equilibrium of the hierarchical game, thereby producing the solution to the original problem. Simulations show the effectiveness of the introduced framework, and demonstrate the superiority of the proposed solution over counterparts.
\end{abstract}
	
}

\maketitle

\IEEEdisplaynontitleabstractindextext

\IEEEpeerreviewmaketitle

\section{Introduction}

\IEEEPARstart{D}{igital} twin (DT) has emerged as a promising technology to create high-fidelity and hyper-realistic virtual counterparts of physical entities, revolutionizing the interaction between physical and digital worlds \cite{BG3}.
With predictive insights and real-time simulations, DT can drive breakthroughs across industries, such as remote healthcare, traffic management, autonomous driving and infrastructure planning \cite{BGour, BG0, BGour2}.
%
To unleash the full potential, the DT construction, which refers to constructing a living and self-evolving visualization model empowered by artificial intelligence (AI) algorithms and deep neural networks \cite{BG1}, is obviously crucial.
This requires not only the collection of real-time sensing data acquired from the ever-changing environment, but also the continuous state updates, to guarantee the synchronization between the DT's virtual model and its corresponding physical entity \cite{Dynamic1}.

Although the DT construction via \textcolor{black}{distributed} sensing has been discussed in the existing work \cite{BGour, BGour2, BG1, Dynamic0, Dynamic1}, all of them were restricted to the centralized framework, where the feature data is collected by deployed sensors, and directly transmitted to a central cloud for processing (including fine-grained model docking, rendering and visualization \cite{VISUAL0, SD3}). However, as the status of physical entity always varies dynamically over the time, such a centralized framework has to be maintained by continuously uploading all data to the central cloud and updating the entire DT model \cite{MC4}, which may incur significant communication and computation overheads. Moreover, the prevalence of data silos in wide-range physical environments may also impede the real-time data collection and utilization \cite{Silos, BGour}, which ultimately degrades the quality of the centrally constructed DT model. To overcome these bottlenecks, we introduce, for the first time, a novel federated DT construction framework under edge-cloud collaboration. In this framework, a DT model to be built at the central cloud can be regarded as a global one being split into and integrating from multiple functional components, i.e., partial-DTs, created at edge servers (ESs). Specifically, i) the central cloud, which is responsible for constructing the DT model, first splits the whole model into multiple partial-DTs; ii) each partial-DT is then assigned to an ES for creation using feature data provided by its associated \textcolor{black}{sensors}; and iii) once all partial-DTs are created, they are forwarded to the central cloud for global integration, i.e., connecting and harmonizing these partial-DTs into a complete global DT model. \textcolor{black}{In practice, this federated DT construction framework can be applied to various practical scenes, such as intelligent transportation and smart factory, where the local server at each intersection or factory creates partial-DTs for road segments or assembly lines using surveillance data collected by traffic or industrial cameras, and then transmits its partial-DT to the headquarter \cite{Rsp1, MC4, zwjBSC}.} By this way, the DT construction is performed with the help of partial-DTs' creations and evolutions in distributed and parallel structures, so that high cost-efficiency and DT model quality can be achieved.

\textcolor{black}{Essentially, the DT model is a complex integration of not only AI models, but also visual models, geometric models, rule models, data analytic models, etc \cite{Rsp1, Rsp2}. While the federated model construction, typically the federated learning, has been extensively studied, the federated DT construction possesses unique features, making it fundamentally different, and thus is worthy to be carefully explored. On one hand, unlike the homogeneous local models trained in the federated learning, which all deal with the same task \cite{FL3, FL4, BG2, FL0, FL1}, partial-DTs in the federated DT construction are heterogeneous, each with distinct model parameters and functions, trained by potentially various sources of feature data. For instance, in the transportation DT system, partial-DTs for vehicles, pedestrians and road scenes can be diverse due to different monitoring areas or sensing equipments, but should be simultaneously created and later integrated as a whole \cite{MC2, MC3}. On the other hand, unlike the federated learning that commonly focuses on training a global model until convergence in a single time frame, the federated DT construction has to be dynamically conducted frame-by-frame in accordance with the DT evolution. For example, the transportation DT should be timely updated to synchronize the real-world road traffics, so that its construction is adaptive rather than static \cite{MC4, Dynamic1}. For exploiting the benefit of the federated DT construction and facilitating its implementation under edge-cloud collaboration, it is required to optimize partial-DT assignments from the cloud to ESs, ES-sensor associations for partial-DT creation, and as well as computation and communication resource allocations for global DT integration. Despite the significance of this issue, it is very challenging for the following reasons.}
\textcolor{black}{\begin{enumerate}
	\item Evidently, under edge-cloud collaboration, DT model splitting, partial-DT assignment, ES-sensor association, partial-DT training and global DT integration are sequential but tightly coupled, meaning that all corresponding resource allocations have to be jointly optimized. Moreover, unlike the federated learning where each local model is statically weighted by a predetermined coefficient, partial-DTs created at ESs may contribute differently to the integration of the global DT model at the cloud \cite{MC0, MC2, TI0}, having different importance weights. Causing by uncertain DT evolutions, such importance weights may be time-varying. This together with unpredictable wireless networking conditions and sensing capabilities, make the optimization uncertain and dynamic, motivating us to design an advanced online algorithm with strong exploration capability for maximizing the system performance in long runs.
	\item In practice, partial-DTs, as multiple functional components of a global DT model, may exhibit certain shared characteristics, such as the same rendering material and kernel functions, implying that the data required for their creations are partly identical \cite{MC0, SD3}. As a result, in our proposed federated DT construction framework, each sensor can simultaneously associate with multiple ESs, participating in different partial-DT creations. This leads to a complicated interrelationship between ESs and sensors, i.e., overlapped ES-sensor associations, necessitating us to integrate the online optimization algorithm by an extension of the analysis and solution in conventional exclusively independent coalitions \cite{NonOverlap, 48} to overlapping coalitions.
\end{enumerate}}

In this paper, to fill the gap in the literature, we propose a novel federated DT construction framework with the support of edge-cloud collaboration.
We aim to maximize the global DT's model quality while minimizing all induced costs, including energy consumption and configuration costs. By taking into account uncertain DT evolutions and resulted impacts, we formulate an online optimization problem for dynamically determining partial-DT assignments, ES-sensor associations, and computation and communication resource allocations.
To circumvent the difficulty in solving this complicated problem, we first transform it into a two-layer hierarchical game. In the upper layer, a two-sided matching subgame is formulated for the cloud to optimize partial-DT assignments. In the lower layer, an overlapping coalition formation game is formulated for ESs to optimize ES-sensor associations and resource allocations. 
After proving an existence of the equilibria within these games, 
we apply the Gale-Shapley (GS) algorithm and particularly develop a switch rules-based overlapping coalition formation (SOCF) algorithm to find the short-term equilibria of upper-layer and lower-layer subgames, respectively. 
Then, we design a deep reinforcement learning (DRL) based solution integrated with stable matching and overlapping coalition formation, called DMO, addressing the dynamic and multi-dimensional decision making problem. The DMO approach captures the time interdependence, and connects the short-term equilibria to a long-term equilibrium of the hierarchical game, providing an efficient solution to the original online optimization problem.

The main contributions of this paper are summarized in the following.
\begin{itemize}
	\item To the best of our knowledge, we are the first to 
	introduce the federated DT construction via \textcolor{black}{distributed} sensing under edge-cloud collaboration. We formulate an online optimization problem that dynamically optimizes partial-DT assignments, ES-sensor associations, and computation and communication resource allocations to maximize the long-term system performance of the considered framework.
	\item We propose a novel game-theoretical solution that first transforms the online optimization problem into a two-layer hierarchical game. Then, based on the equilibrium analyses, we leverage the GS algorithm and develop an SOCF algorithm to efficiently obtain the short-term equilibria of upper and lower subgames. On top of this, we further design the DMO approach to well accommodate dynamic settings.
	\item We conduct extensive simulations to evaluate the performance of the proposed federated DT construction framework along with the game-theoretic optimization solution. Results show the effectiveness of the framework and the superiority of the solution in terms of increasing the gain while reducing the cost in the DT construction compared to counterparts.
\end{itemize}

The rest of this paper is organized as follows: Section \ref{RW} reviews the related work and highlights the novelties of this paper. Section \ref{SM} presents the system model of the federated DT construction framework and formulates the corresponding optimization problem. In Section \ref{Solu}, a game-theoretic solution approach with two-sided matching and overlapping coalitions is proposed and analyzed. Simulation results are given in Section \ref{SR}, followed by the conclusion in Section \ref{CO}.

\section{Related Work}\label{RW}
As a key aspect to the promotion of DT applications, the DT construction has drawn dramatically increasing attentions from both academia and industry. 
For example, Microsoft released the Azure DT platform \cite{Industry0} which allowed users to create various DT models on the cloud. A startup company called Carbontribe, offered an AI-driven DT prototype \cite{Industry1}  that can accurately predict the greenhouse gas emissions using satellite and biodiversity data. 
Besides, the DT construction over networks, enabled by edge computing, \textcolor{black}{distributed} sensing, etc. has also be recently studied.
For instance, Yang et al. in \cite{Dynamic1} explored a vehicular DT construction framework, in which the traffic DT was built at the central cloud and dynamically synchronized with the moving vehicles, assisted by the relay of roadside units.
Li et al. in \cite{BG1} proposed a continual learning-driven DT construction framework to build and update the DT on multiple ESs using the data collected from Internet-of-Things devices, aiming to maximize the long-term model accuracy. 
Yang et al. in \cite{BGour2, yyy2} developed a personalized DT construction framework, where DTs with both generic and customized parts were dynamically deployed on local ESs under the control of a remote cloud, utilizing the collected feature data from nearby sensors.
However, all these work focused on the centralized DT construction framework, ignoring the potential of constructing the DT in a federated manner.

Some researchers have investigated the federated model construction, particularly the federated learning and its variants.  
Name a few, Zhang et al. in \cite{FL3} studied the intelligent reflecting surfaces-assisted federated learning, where identical models were trained on local clients, targeting to minimize the energy consumption till the convergence of model accuracy.
Pham et al. in \cite{FL4} developed an air-ground federated learning structure, where multiple ground clients were hired to first train sub-models for the same task, and the global model was then aggregated at a drone server.
Further taking into account the system dynamics, Sun et al. in \cite{BG2} devised a Stackelberg game-empowered federated learning framework, where clients were asked to perform the same task and were dynamically selected via an incentive mechanism, assisted by a DT of the whole air-ground network. 
However, the solutions in all these work can hardly be employed for the DT construction because unlike general models, the DT should be dynamically evolving with the ever-changing physical environments and is composed by heterogeneous partial-DTs, each of which contributes differently to the global DT.


Some preliminary efforts have been devoted to studying the online optimization of task splitting and assignments, client associations, and resource allocations, that may possibly applied for the federated DT construction. For example, Kobayashi et al. in \cite{DynamicAA} formulated a dynamic combinatorial optimization problem for jointly optimize the channel assignment and base station association for downlink-uplink decoupling in wireless networks.
Ng et al. in \cite{NonOverlap} proposed a reputation-aware hedonic coalition formation game approach to divide the clients into multiple non-overlapping groups for associating with an intermediate aggregator within a hierarchical federated learning framework.
Xu et al. in \cite{FL0} designed an online learning algorithm to dynamically and jointly select a non-overlapping subset of clients associated with one aggregator, which was then assigned to perform one federated learning request, aiming to maximize the long-term energy efficiency. Nevertheless, all these work assumed that tasks were independent with each other, so that client associations were exclusive without any overlaps. This does not fit the federated DT construction framework under edge-cloud collaboration, where each sensor may be associated with multiple ESs simultaneously, participating in various partial-DT creations.

In summary, this work differs from the existing literature in the following aspects: i) we introduce a federated DT construction framework supported by edge-cloud collaboration to achieve higher model quality and lower costs in a distributed manner, rather than the traditional centralized ones; ii) we consider the uncertain DT evolution, along with the heterogeneity of partial-DTs, and formulate an online optimization problem to jointly and dynamically determine partial-DT assignments, ES-sensor associations and resource allocations; iii) we study overlapping coalitions in ES-sensor associations for partial-DT creation, instead of the conventional exclusively independent coalitions, and make them accommodate the dynamic settings. 

\section{System Model and Problem Description}\label{SM}
In this section, we first present a system overview on how a DT model is constructed via \textcolor{black}{distributed} sensing and evolved in a federated manner. Then, we describe the partial-DT creation on local ESs and global DT integration on the central cloud. After that, aiming to create a high-quality DT model while reducing the costs from data sensing, partial-DT creation and global DT integration in long runs, we formulate an online optimization problem.

\subsection{System Overview}
\begin{figure}[!t]
	\centering
	\includegraphics[width=\columnwidth]{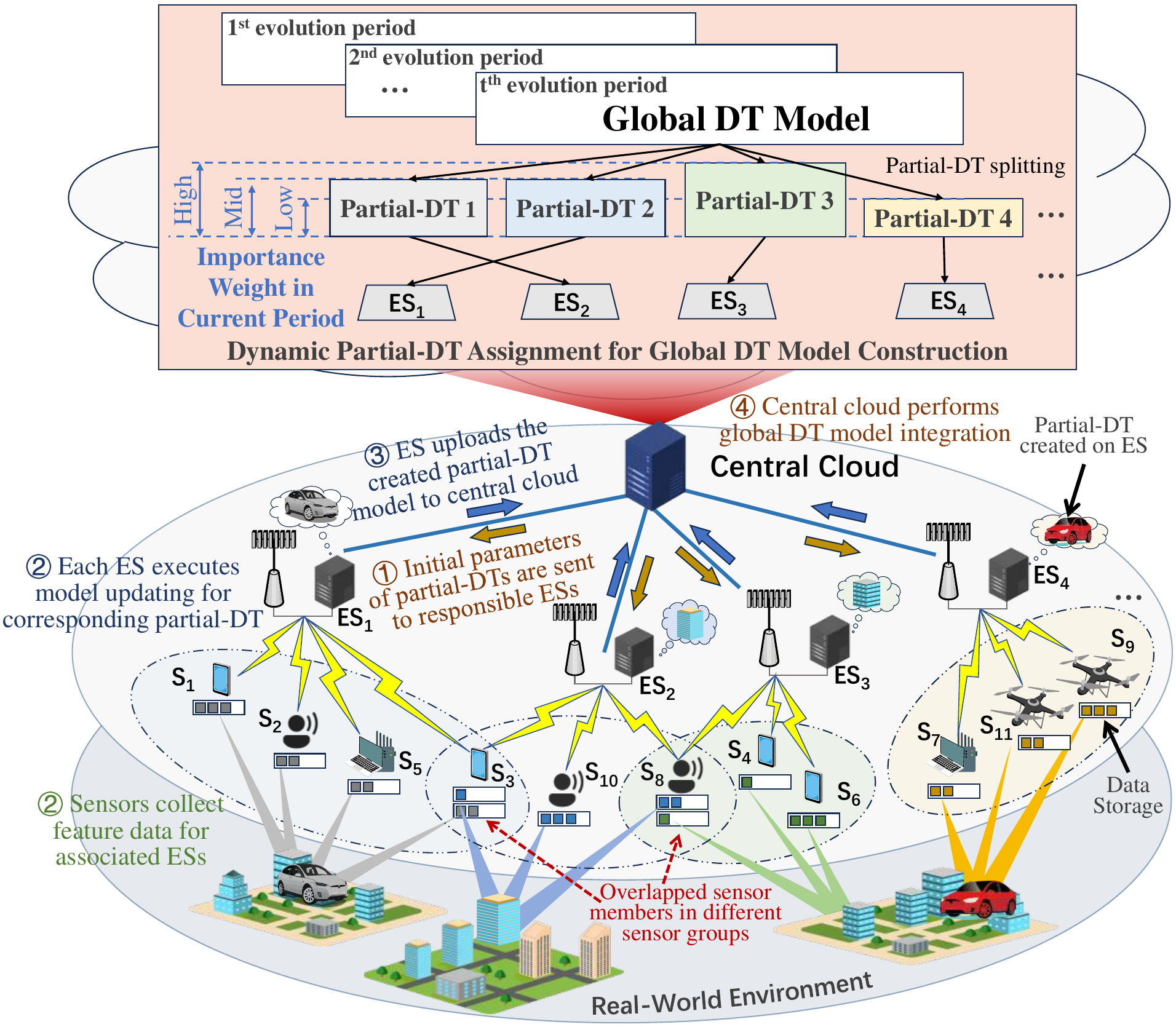}\\
	\caption{An illustration of the federated DT construction by edge-cloud collaboration. In this framework, the cloud first splits the global DT model into partial-DTs, and then each ES respectively creates its assigned partial-DT using data collected by its associated sensors. Finally all partial-DTs are uploaded to the cloud for global DT integration.}\label{systemmodel}
	\vspace{-1.2em}
\end{figure}
Consider a federated DT construction framework building upon an edge-cloud collaborative system, as shown in Fig. \ref{systemmodel}, consisting of a central cloud $CS$, a set of geographically dispersed local ESs $\mathcal B$ with $|\mathcal B| = B$, and a set of randomly scattered sensors $\mathcal N$ with $|\mathcal N| = N$. In this framework, the DT model is seen as a global one, which can be split into multiple functional components \cite{MC0, SD3, MC2}, i.e., partial-DTs, which are simultaneously created on ESs. These partial-DTs are essentially different partial digital replicas of the physical entity, and can be obtained by using the featured data collected from selected groups of associated sensors \cite{MC4, SD2, SD3}. Then, integrating all partial-DTs from ESs to the central cloud, a global DT model can be constructed. Following the DT configurations \cite{BG1, BG2}, we can define that
\begin{equation}
	\begin{aligned}
		Global\mbox{-}DT=\big \langle \{Partial\mbox{-}DT ~c\}_{\forall c \in \mathcal{C}} \big \rangle,
	\end{aligned}
\end{equation}
where $\mathcal{C}$ denotes the set of all partial-DTs of the global DT model, $c\in \mathcal{C}$ indicates one specific partial-DT, and $\langle \cdot \rangle$ represents the integration of all partial-DTs.

Although each partial-DT $c\in \mathcal{C}$ is created in a distributed manner on a specific ES, the central cloud has to first inform each ES by the initial values of core parameters in partial-DT $c$ (e.g., model structures, rendering requirements, configuration information \cite{MC0, SD2}), performing the DT model splitting and partial-DT assignments. Then, sensors are divided into multiple groups, each group is associated with one ES for feature data collection. Here, since different ESs may require the same data for creating partial-DTs with shared characteristics, we do not restrict one sensor to be associated with at most one ES, implying that each of them can be potentially in multiple groups, simultaneously helping different ESs' partial-DT creations \cite{OCFG0, SO0}. 
By utilizing the initial values informed by the cloud and the feature data acquired by associated sensors in the physical-virtual mapping process, each ES can then establish a partial-DT model, for global DT integration.

Note that DT model should be timely constructed and evolved to maintain a high quality in long runs because its original goal is to replicate realistically the physical world which always vary dynamically \cite{Dynamic0}.
Consequently, the proposed federated DT construction procedure should evolve over the time, in which all partial-DTs has to be continuously recreated and updated by time-increasing data collected from sensors in real time. To capture the system dynamics stem from long-term evolution nature of the DT, we employ a time-framed system, where each time frame
$t\in\{1,2,...,T\}$ refers to the duration of one DT evolution period. Furthermore, in practice, partial-DTs may contribute differently to the integration of the global DT model, meaning that they have different importance weights \cite{TI0}, and these weights can also vary with DT evolutions. Thus, we introduce a dynamic importance weight index, defined as $\mathcal{I}=\{I_c(t)\}_{c\in \mathcal C}$, where $I_c(t)\in[0,1]$ is the importance weight of partial-DT $c \in \mathcal{C}$ in time frame $t$, which can be revealed by the specific application requirements of each partial-DT.

Overall, we aim to optimize the long-term system performance of the federated DT construction framework, i.e., enhancing the quality of the global DT model while reducing the costs from data sensing, partial-DT creation and global DT integration in long term.
Particularly, in each time frame $t$, we are required to dynamically determine i) which partial-DT should be assigned to which ES for creation, so that partial-DTs with larger importance weights can always be handled by ESs with both more computation resources and more associated sensors in feature data collection; and ii) which sensors should be grouped to associate with which ES for assisting its partial-DT creation, allowing that one sensor can be associated with multiple ESs, forming overlapping groups. For convenience, the main notations used in this paper are listed in Table 1.

\begin{table}[!t]
	\footnotesize
	\caption{IMPORTANT NOTATIONS IN THIS PAPER}
	\vspace{-0.6em}
	\renewcommand{\arraystretch}{1.1}
	\label{tab2}
	\centering
		\begin{tabular}{l|l}
			\hline
			\bfseries Symbol&\bfseries Explanation\\ 
			\hline
			$A_{c,b}(t)$&model quality of partial-DT $c$ created on ES $b$\\
			$A_c^{Req}$ &required maximum model accuracy of partial-DT $c$\\
			$A^{Global}(t)$&quality of the global DT model\\		
			$C^{Conf}$&potential configuration cost\\	
			$Co_b(t)$&all sensors associated to ES $b$\\
			$\mathcal{C},\mathcal{B},\mathcal{N}$&set of partial-DTs, ESs or sensors \\
			$\mathcal{CP}(t)$ & coalition partition\\
			$d_c(t)$&amount of data collected for partial-DT $c$\\
			$E^{Total}(t)$&total energy consumption\\	
			$\varPhi_t(i)$ & bidirectional matching\\
			$\mathcal G^{H}$&hierarchical game \\
			$\mathcal G^{Up}_{t},\mathcal G^{Low}_{t}$&upper-layer and lower-layer subgames\\
			$I_c(t)$&importance weight of partial-DT $c$\\
			$P_n$&transmit power of sensor $n$\\
			$\pi^R_b(t)$ & resource allocation for ES $b$\\			
			$R^T_{n,b}(t)$&wireless transmission rate from sensor $n$ to ES $b$\\
			$T_b(t)$&decision on training rounds of ES $b$\\
			$\tau^{Total}(t)$&total latency\\		
			$\mathcal{U}(t), \mathcal{U}_{Sys}$&short-term and long-term system performance\\
			$x_{c,b}(t)$&decision on partial-DT assignments\\		
			$y_{b,n}(t)$&decision on ES-sensor associations\\			
			$z_{b,n,w}(t)$&decision on subcarrier allocations\\
			\hline
		\end{tabular}
\end{table}

\subsection{Model Quality of Federated DT Construction}
In the federated DT construction framework, the quality of the global DT model constructed at the central cloud within each time frame $t$, denoted by $A^{Global}(t)$, depends on the partial-DT assignments, time-varying importance weight of each partial-DT and achieved model quality of each partial-DT created by the responsible ES, which can be expressed as
\begin{equation}
	\begin{aligned}
		A^{Global}(t)=&\sum\nolimits_{c\in \mathcal C} \sum\nolimits_{b\in \mathcal B} x_{c,b}(t) I_c(t) A_{c,b}(t), 
	\end{aligned}
\end{equation}
where $I_c(t)$ stands for the importance weight of each partial-DT $c\in \mathcal{C}$ in time frame $t$, $A_{c,b}(t)$ represents the model quality of partial-DT $c\in \mathcal{C}$ created by ES $b \in \mathcal{B}$ in time frame $t$, and $x_{c,b}(t) \in \{0,1\}$ is the decision on partial-DT assignment, indicating whether partial-DT $c \in \mathcal{C}$ is assigned to be created on ES $b \in \mathcal{B}$ in time frame $t$. Note that we impose the following constraints
\begin{equation}
	\begin{aligned}
		\hspace{-1.5mm}\sum\nolimits_{b\in \mathcal B} x_{c,b}(t) \hspace{-0.5mm}= \hspace{-0.5mm} 1 , \forall c \hspace{-0.5mm} \in  \hspace{-0.5mm}\mathcal C \text{, and } \sum\nolimits_{c\in \mathcal C} x_{c,b}(t) \hspace{-0.5mm}\leq  \hspace{-0.5mm}1 , \forall b  \hspace{-0.5mm}\in \hspace{-0.5mm} \mathcal B, \label{zcb}
	\end{aligned}
\end{equation}
meaning that all partial-DTs have to be created for guaranteeing the global DT model integrity, and one ES is allowed to handle the creation of at most one partial-DT. 

To further measure $A_{c,b}(t)$, we extend the well-known definition of model accuracy in the federated learning \cite{FL0, FL1, FL3, FL4} as the performance of partial-DT creations.
\textcolor{black}{Namely, $A_{c,b}(t)$ is calculated based on both the amount of cumulatively collected feature data and the training rounds that ES $b\in \mathcal{B}$ determines to take for creating partial-DT $c\in \mathcal{C}$ under iteratively model training using the feature data\footnote{Note that, the model quality of partial-DT is approximated by a weighted model accuracy with respect to the data size, and it is formulated in this particular way for mathematical tractability, while our proposed solution can adapt to other customized model quality measurement tailored to specific DT applications.}. Mathematically, $A_{c,b}(t)$ can be written as
\begin{equation}
	\begin{aligned}
		A_{c,b}(t) =& x_{c,b}(t) \min \big \{ A_c^{Req}, \varGamma \big (\sum\nolimits_{i=1}^t d_c(i) \big) \\
		&\cdot \big ( 1 - 2^{\left(-T_b(t)(2-L\delta)\delta \gamma \right)/2} \big ) \big\} , \label{Accu}
	\end{aligned}
\end{equation}
where $A_c^{Req}$ stands for the desired maximum model quality of partial-DT $c \in \mathcal{C}$, $\varGamma \big (\sum\nolimits_{i=1}^t d_c(i) \big)  \big ( 1 - 2^{\left(-T_b(t)(2-L\delta)\delta \gamma \right)/2} \big )$ indicates the model quality that each partial-DT $c\in \mathcal{C}$ can achieve after training multiple rounds with the cumulatively collected data \cite{FL0, FL1}. Particularly, $\varGamma \big (\sum\nolimits_{i=1}^t d_c(i) \big)= \big( \log_2(\frac{\sum\nolimits_{i=1}^t d_c(i)}{t\beta}+1) \big)^2$ represents the impact from the cumulatively collected feature data following \cite{BGour2, Rsp33}, expressed as a discount coefficient calculated by logarithmic normalization function with parameter $\beta$ \cite{LogNormal0}, and $\sum\nolimits_{i=1}^t d_c(i)$ denotes the amount of cumulatively collected feature data for partial-DT $c \in \mathcal{C}$ up to time frame $t$. $1 - 2^{\left(-T_b(t)(2-L\delta)\delta \gamma \right)/2}$ calculates an approximate model accuracy in terms of the logistic loss function \cite{FL3, FL0, FL1}, where $T_b(t)$ is the number of training rounds determined to be taken by ES $b \in \mathcal{B}$ in time frame $t$, $L$, $\gamma$ and $\delta$ are pre-defined hyperparameters, indicating that the loss function in training the AI model for DT satisfies $L$-Lipschitz continuous and $\gamma$-strongly convex, with $\delta \in (0, 2/L)$ being the corresponding learning rate \cite{FL1}.
}

Observed from (\ref{Accu}) that $A_{c,b}(t)$ increases with $T_b(t)$ but its marginal benefit decreases when $T_b(t)$ is considerably large. To be cost-efficient, $T_b(t)$ is naturally constrained as
\begin{equation}
	\begin{aligned}
		T_b(t) \leq T^*_b(t), \forall b \in \mathcal{B}, \label{Tbt}
	\end{aligned}
\end{equation}
where $T^*_b(t)$ is the upper bound of $T_b(t)$, which can be calculated by letting $A_{c,b}(t) = A^{Req}_c$ in (\ref{Accu}), i.e.,
\begin{equation}
	\begin{aligned}
		\hspace{-2.2mm}T^*_b(t) \hspace{-0.7mm} = \hspace{-0.7mm} \sum_{c\in \mathcal C} x_{c,b}(t) \hspace{-0.9mm}\left \lceil \hspace{-0.7mm}  \frac{2 }{(L\delta \hspace{-0.7mm} - \hspace{-0.7mm} 2)\delta \gamma} \log_2(1 \hspace{-0.5mm}-\hspace{-0.5mm} \frac{A_c^{Req}}{ \varGamma \big (\sum\nolimits_{i=1}^t \hspace{-0.5mm} d_c(i) \big)}) \hspace{-0.7mm} \right \rceil \hspace{-0.5mm}.\hspace{-0.9mm}
	\end{aligned}\label{T*BT}
\end{equation}

Furthermore, considering that the ES-sensor association plays a vital role in feature data collection for partial-DT creations, the total amount of new data collected in each time frame $t$ for partial-DT $c \in \mathcal{C}$ is expressed as
\begin{equation}
	\begin{aligned}
		d_c(t) = \sum\nolimits_{b \in \mathcal B}x_{c,b}(t)\sum\nolimits_{\forall n \in \mathcal N}y_{b,n}(t) d_{n,c}(t),
	\end{aligned}
\end{equation}
where $d_{n,c}(t)$ stands for the amount of feature data that each sensor $n$ can obtain for partial-DT $c\in \mathcal{C}$ in time frame $t$, and $y_{b,n}(t) \in \{0,1\}$ is the decision on the ES-sensor association, denoting whether each sensor $n \in \mathcal{N}$ is determined to be associated with ES $b\in \mathcal{B}$ in time frame $t$. Recall that we allow each sensor to be associated with multiple ESs, while the total number of ESs that one sensor can be associated with should be restricted due to the hardware limitations of the multi-cast transmissions \cite{OCFG1, OCFG3}, i.e.,
\begin{equation}
	\begin{aligned}
		\sum\nolimits_{b \in \mathcal B} y_{b,n}(t) \leq L_n, \forall n \in \mathcal{N}, \label{sensorassociation} \\
	\end{aligned}
\end{equation}
where $L_n$ is the maximum number of ESs that each sensor $n \in \mathcal{N}$ can be associated with.

\subsection{Latency and Energy Consumption in Federated DT Construction}
The total latency of federated DT construction in each time frame $t$ is the sum of i) the latency in preparing all partial-DTs at ESs, which includes sensing data transmission latency $\tau^{Dtr}_b(t)$, historical data migration\footnote{\textcolor{black}{Note that the historical data is always essential as the foundation for creating each partial-DT because it cannot be built from scratch with real-time collected sensing data only \cite{HisData,BG0}. For example, in transportation DT system, historical data such as past traffic patterns and accident history provides the fundamental knowledge required for traffic prediction and also mirrors the past states of the DT model, which is necessary for accurate model construction.}} latency $\tau^{Back}_b(t)$, partial-DT creation latency $\tau^{Cre}_b(t)$ and partial-DT uploading latency $\tau^{Mtr}_{b}(t)$; and ii) the global DT integration latency $\tau^{I}_{CS}(t)$ at the cloud, i.e.,
\begin{align}
	\tau^{Total}(t) =& \max_{b\in \mathcal B} \{\tau^{Dtr}_b(t) + \tau^{Back}_b(t) +\tau^{Cre}_b(t) + \tau^{Mtr}_{b}(t) \} \notag \\
 &+ \tau^{I}_{CS}(t), \label{TotalLatency}
\end{align}
where $\max_{b\in \mathcal B} \{\cdot\}$ calculates the maximum latency among all partial-DTs on all ESs.
Hence, the total energy consumption of federated DT construction in each time frame $t$ is
\begin{align}
	E^{Total}(t) =& \sum\nolimits_{b\in \mathcal{B}} \big ( E^{Dtr}_{b}(t) + E^{Back}_{b}(t) + E^{Cre}_b(t) \notag \\
	& + E^{Mtr}_{b}(t) \big ) + E^I_{CS}(t), \label{TotalEnergy}
\end{align}
where $E^{Dtr}_{b}(t)$, $E^{Back}_{b}(t)$, $E^{Mtr}_{b}(t)$ and $E^{Cre}_b(t)$ respectively denote the energy consumption of sensing data transmission, historical data migration, partial-DT creation and partial-DT uploading in preparing each partial-DT $c \in \mathcal{C}$ on its responsible ES $b \in \mathcal{B}$, and $E^I_{CS}(t)$ represents the energy consumption of global DT integration at the cloud, all in time frame $t$. 

Specifically, $\tau^{Dtr}_b(t)$ and $E^{Dtr}_{b}(t)$ are the maximum latency and total energy consumption of all sensors associated with each ES $b \in \mathcal{B}$, respectively, in collecting data for partial-DT $c \in \mathcal{C}$, i.e.,
\begin{equation}
	\begin{aligned}
		\tau^{Dtr}_b(t) = \max_{n\in \mathcal N} \{ x_{c,b}(t)y_{b,n}(t) d_{n,c}(t) \big / {R^T_{n,b}(t)} \}, \label{tauDtr}
	\end{aligned}
\end{equation}
\begin{equation}
	\begin{aligned}
		E^{Dtr}_{b}(t)=\sum\nolimits_{n\in\mathcal{N}} y_{b,n}(t) P_n x_{c,b}(t) d_{n,c}(t) \big / {R^T_{n,b}(t)},
	\end{aligned}
\end{equation}
where $x_{c,b}(t) y_{b,n}(t) d_{n,c}(t)$ indicates the amount of sensing data acquired by each sensor $n \in \mathcal{N}$ associated with ES $b \in \mathcal{B}$ assigned to create partial-DT $c\in \mathcal{C}$, $P_n$ denotes the transmit power of sensor $n \in \mathcal{N}$, and $R^T_{n,b}(t)$ stands for the wireless transmission rate from sensor $n \in \mathcal{N}$ to ES $b \in \mathcal{B}$ in time frame $t$.

Particularly, $R^T_{n,b}(t)$ can be calculated as
\begin{equation}
	\begin{aligned}
		&\hspace{-2.0mm}R^T_{n,b}(t) = y_{b,n}(t) z_{b,n,w}(t) W \log_2 \Big ( 1 +\\
		&~~\frac{P_{n} H_{n,b,w}(t)}{\big (\sum_{b'\hspace{-0.5mm}\in \mathcal B}\sum_{n'\hspace{-0.5mm}\in \mathcal N} y_{b'\hspace{-0.5mm},n'}(t) z_{b'\hspace{-0.5mm},n'\hspace{-0.5mm},w}(t)P_{n'} H_{n'\hspace{-0.5mm},b,w}(t)\hspace{-0.5mm}+\hspace{-0.5mm}n_{b}^2\big )}  \Big ),\hspace{-2.0mm}
	\end{aligned}
\end{equation}
where $W$ represents the bandwidth of each subcarrier $w \in \mathcal{W}$, $H_{n,b,w}(t)$ is the instantaneous channel gain from wireless sensor $n \in \mathcal{N}$ to ES $b \in \mathcal{B}$ on subcarrier $w \in \mathcal{W}$, $\sum_{b'\in \mathcal B}\sum_{n'\in \mathcal N}y_{b',n'}(t) z_{b',n',w}(t)P_{n'} H_{n',b,w}(t)$ indicates the co-channel interference at ES $b\in \mathcal{B}$, $n_{b}^2$ signifies the additive white Gaussian noise at ES $b\in \mathcal{B}$, and $z_{b,n,w}(t) \in \{0,1\}$ is the decision on subcarrier allocation \cite{OCFG1}, denoting whether each subcarrier $w \in \mathcal{W}$ is allocated to sensor $n \in \mathcal{N}$ that is associated with ES $b\in \mathcal{B}$ in time frame $t$. Obviously, we should have
\begin{equation}
	\begin{aligned}
		\sum\nolimits_{n \in \mathcal N} \sum\nolimits_{w \in \mathcal{W}} y_{b,n}(t) z_{b,n,w}(t) \leq |\mathcal{W}|,  \label{PCB8}
	\end{aligned}
\end{equation}
where $|\mathcal{W}|$ is the maximum number of subcarriers that one ES can provide to all its associated sensors.

Besides the newly collected sensing data, historical data is also indispensable in the partial-DT creation. For each partial-DT $c\in \mathcal{C}$ created on ES $b\in \mathcal{B}$, this data should be dynamically migrated from the previously assigned ES $b'\in \mathcal{B}$ in time frame $t-1$ to ES $b$ in time frame $t$. Thus, we have
\begin{align}
	\tau^{Back}_{b}(t) &=  \sum\nolimits_{b' \in \mathcal{B}}x_{c,b'}(t-1) \sum\nolimits_{i=1}^{t-1}d_c(i)\big/{R^T_{b',b}},\\
	E^{Back}_{b}(t) &=  \sum\nolimits_{b' \in \mathcal{B}} x_{c,b'}(t-1) P_{b'} \sum\nolimits_{i=1}^{t-1}d_c(i)\big/{R^T_{b',b}},
\end{align}
where $\sum\nolimits_{i=1}^{t-1}d_c(i)$ indicates the amount of cumulatively historical data up to time frame $t-1$, $R^T_{b',b}$ stands for the transmission rate from ES $b'\in \mathcal{B}$ to another ES $b\in \mathcal{B}$ via the fiber link \cite{BGour2}, and $P_{b'}$ is the transmit power of ES $b'$.

 
In each time frame $t$, since creating the assigned partial-DT $c \in \mathcal{C}$ on ES $b \in \mathcal{B}$ involves the model training with multiple rounds $T_b(t)$ for improving the model quality, the latency and energy consumption for such partial-DT creation can be respectively written as
\begin{align}
	\tau^{Cre}_{b}(t)&=T_b(t) \sum\nolimits_{n\in \mathcal N}{\big (y_{b,n}(t)x_{c,b}(t) d_{n,c}(t)\Upsilon_b \big )}/{F_{b}},\\
	E^{Cre}_b(t)&=\rho_b(F_b)^2 T_{b}(t) \Upsilon_b,
\end{align}
where $\Upsilon_b$ stands for the CPU cycles required for computing one byte of data on ES $b \in \mathcal{B}$ (measured by s/byte) \cite{FL1}, $F_{b}$ is the CPU speed of ES $b \in \mathcal{B}$ (measured by cycles/s), and $\rho_b$ denotes the effective switched capacitance of ES $b \in \mathcal{B}$.

After the partial-DT creation, each ES $b\in \mathcal{B}$ will upload its created partial-DT $c\in \mathcal{C}$ to the cloud. The corresponding latency and energy consumption can be respectively expressed as
\begin{align}
	\tau^{Mtr}_{b}(t)&= x_{c,b}(t)D_c \big / R^T_{b, CS}, \label{tauTbt}\\
	E^{Mtr}_{b}(t)&= x_{c,b}(t)P_{b}D_c \big / R^T_{CS,b}, \label{EMtr}
\end{align}
where $D_c$ is the model size of each partial-DT $c \in \mathcal{C}$, and $R^T_{b, CS}$ denotes the transmission rate from ES $b\in \mathcal{B}$ to the cloud via the fiber link \cite{BGour2}. 

Furthermore, the latency for global DT integration in each time slot $t$ is caused by processing all partial-DTs aggregated at the cloud, i.e.,
\begin{align}
	\tau^{I}_{CS}(t) = \sum\nolimits_{c\in \mathcal{C}}\sum\nolimits_{b\in \mathcal{B}} x_{c,b}(t)D_c\Upsilon_{CS} \big  / F_{CS},
\end{align}
where $\Upsilon_{CS}$ is the number of CPU cycles required for integrating a single byte data of partial-DT models, and $F_{CS}$ is the CPU speed of the cloud.
The energy consumption $E^{I}_{CS}(t)$ of global DT integration due to executing resource-intensive instructions over both CPU and GPU resources on the cloud (e.g., fine-grained model docking, rendering and visualization \cite{VISUAL0, SD3}),
following \cite{FL0}, $E^{I}_{CS}(t)$ can be calculated as
\begin{equation}
	\begin{aligned}
		E^{I}_{CS}(t) =& \tau^{I}_{CS}(t) \big (\sum\nolimits_{c \in \mathcal{C}} \sum\nolimits_{b\in \mathcal{B}} x_{c,b}(t) \frac{\zeta_{CS} H_{CS,c}}{\tau^{I}_{CS}(t)}   \\
		&\cdot P^{Max}_{CS}+P^{Idle}_{CS} + P^{Leak}_{CS} \big ),
	\end{aligned}
\end{equation}
where $\zeta_{CS}$ is the access rate of the cloud's single processing unit, $H_{CS,c}$ is the average number of instructions in integrating partial-DT $c \in \mathcal{C}$, $P^{Max}_{CS}$, $P^{Idle}_{CS}$ and $P^{Leak}_{CS}$ are hardware parameters related to the cloud's processing units \cite{FL0}.

\subsection{Problem Formulation}
To evaluate the performance of the federated DT construction framework, for each single time frame $t$, we take the difference between i) the gain of the global DT model quality $A^{Global}(t)$, and ii) the cost of total energy consumption $E^{Total}(t)$, as
\begin{equation}
	\begin{aligned}
		\mathcal{U}(t) = \xi A^{Global}(t)-\kappa E^{Total}(t),
	\end{aligned}
\end{equation}
where $\xi$ and $\kappa$ are weight coefficients.

Furthermore, when expanding $\mathcal{U}(t)$ over multiple time frames for the long-term optimization, partial-DT assignments among ESs can be dynamically adjusted, and triggered by this, ES-sensor associations may also be dynamically varied. This incurs an additional configuration cost between any two consecutive time frames, e.g., signaling overheads caused by notifying such changes \cite{myICC}, \cite{myWCNC}. Considering this, we define the long-term system performance of the federated DT construction framework as
\begin{equation}
	\begin{aligned}
		\mathcal{U}_{Sys} = \lim_{T\rightarrow \infty} \frac{1}{T}\sum\nolimits_{t = 0}^{T}\mathcal{U}(t) - C^{Conf}\mathcal{F}(t),
	\end{aligned}
\end{equation}
where $C^{Conf}$ is a pre-known cost coefficient, 
and $\mathcal{F}(t)=\sum_{b\in \mathcal{B}}\sum_{n\in \mathcal{N}} y_{b,n}(t) \oplus y_{b,n}(t-1)$ indicates the total number of changes of ES-sensor associations in time frame $t$, with $\oplus$ being the exclusive OR operator.

Then, with the objective of maximizing $\mathcal{U}_{Sys}$, we formulate a long-term optimization problem to jointly and dynamically determine partial-DT assignments $x_{c,b}(t)$, ES-sensor associations $y_{b,n}(t)$, subcarrier allocations $z_{b,n,w}(t)$ and training rounds $T_b(t)$ in every time frame $t$, denoted as $\Pi_{Sys}$ $=\big\{x_{c,b}(t), y_{b,n}(t), z_{b,n,w}(t), T_b(t)\big\}_{\forall t,c,b,w}$, which is given bellow.
\begin{subequations}
\begin{align}  
	[\mathcal{P}_{1}]:~&\max_{\Pi_{Sys}}~ \mathcal{U}_{Sys} \label{OF}\\
	s.t. , ~ & ~(\ref{zcb}), ~(\ref{sensorassociation}), ~(\ref{PCB8}),\notag \\
	&\tau^{Total}(t) \leq \delta_T, \label{RealTime}\\
	& R^T_{n,b}(t)\geq R^T_{Min}, \forall b \in \mathcal{B}, \forall n \in \mathcal{N}, \label{ESaccess}\\
    &x_{c,b}(t)\in \{0,1\},~\forall c \in \mathcal C, \forall b \in \mathcal B, \label{PC3}\\
	&y_{b,n}(t)\in \{0,1\},~\forall b \in \mathcal B, \forall n \in \mathcal N, \label{PC1}\\
	&z_{b,n,w}(t)\in \{0,1\},~\forall b \in \mathcal B, \forall n \in \mathcal N, \forall w \in \mathcal{W}, \label{PC2}\\
	&T_b(t) \leq T^*_b(t), \forall b \in \mathcal{B}, \label{PC4}
\end{align}
\end{subequations}
where constraint (\ref{RealTime}) enforces that the cloud should integrate all partial-DTs for model integrity within $\delta_T$, which stands for the time duration of one DT evolution period; constraint (\ref{ESaccess}) ensures that $R^T_{n,b}(t)$ should not be smaller than a minimum value $R^T_{Min}$ for stable communications between sensors and ESs, and constraints (\ref{PC3}) - (\ref{PC4}) are the defined ranges of all decision variables.


\textcolor{black}{\textbf{Remark.} Clearly, solving problem $[\mathcal{P}_{1}]$ directly is very challenging because i) $[\mathcal{P}_{1}]$ is an integer programming problem with a non-convex objective function and multi-dimensional decision variables, leading to the potential curse of dimensionality; ii) since both the importance weights of partial-DTs in the federated DT construction and partial-DT preparations at ESs are uncertain and time-varying, it is necessary to develop an online algorithm with the strong exploration capability; iii) each sensor can be associated with multiple ESs simultaneously, making the traditional ES-sensor association methods, such as the merge-and-split algorithm and Shapley value-based approaches, inapplicable, as they all assume non-overlapping settings; and iv) the configuration cost incurred from the dynamic changes of ES-sensor associations for partial-DT creations increases the interdependence of the decisions across different time frames, meaning that the ES-sensor relationship becomes even more complicated in long runs.	
}

\section{A Game-Theoretic Approach with Two-sided Matching and Overlapping Coalitions}\label{Solu}
In this section, we first transform the original problem $[\mathcal{P}_{1}]$ into a more tractable two-layer hierarchical game consisting of a two-sided matching game and an overlapping coalition formation game. Then, we analyze these games in detail and propose a DRL-based solution to obtain the corresponding equilibrium in an online manner.
\subsection{Problem Transformation to a Hierarchical Game}

%


We observe from the problem formulation in (\ref{OF})-(\ref{PC4}) that solving $[\mathcal{P}_{1}]$ is equivalent to simultaneously solve the following two problems, i.e., 
\begin{align}  
	[\mathcal{P}_{2}]:~&\max_{\{x_{c,b}(t)\}_{\forall t}}~ \lim_{T\rightarrow \infty} \frac{1}{T} \sum\nolimits_{t = 0}^{T} U_{CS}(t) \label{P2}\\
	s.t. , ~ & ~(\ref{zcb}), ~(\ref{RealTime}), ~(\ref{PC3}).\notag \\
	[\mathcal{P}_{3}]:~&\hspace{-2.4mm}\max_{\{y_{b\hspace{-0.2mm},n}(t), z_{b\hspace{-0.2mm},n\hspace{-0.2mm},w}(t), T_b(t)\}_{\forall t}}~ \hspace{-1.65mm}\lim_{T\rightarrow \infty}\hspace{-0.5mm} \frac{1}{T}\hspace{-0.5mm}\sum\nolimits_{t = 0}^{T} \sum\nolimits_{b 
		\in \mathcal{B}} \hspace{-0.5mm}U_{b}(t) \label{P3}\\
	s.t. , ~ & ~(\ref{sensorassociation}), ~(\ref{PCB8}), ~(\ref{RealTime}), ~(\ref{ESaccess}), ~(\ref{PC1}), ~(\ref{PC2}), ~(\ref{PC4}).\notag 
\end{align}
Here, $[\mathcal{P}_{2}]$ and $[\mathcal{P}_{3}]$ can be seen as optimization problems of the cloud and each ES, respectively, including decision variables only related to themselves. While these two problems share a common term in objective functions and the same constraint (\ref{RealTime}), so that their decisions are correlated. To be more specific, in problem $[\mathcal{P}_{2}]$, $U_{CS}(t) \hspace{-0.6mm}=\hspace{-0.5mm}  \sum\nolimits_{b\in \mathcal B} \xi\sum\nolimits_{c\in \mathcal C}  x_{c,b}(t) I_c(t) A_{c,b}(t)\hspace{-0.5mm}-\hspace{-0.5mm}\kappa E^{Total}(t)$ calculates the utility of the cloud in each time frame $t$. In problem $[\mathcal{P}_{3}]$, $U_{b}(t) \hspace{-0.6mm}=\hspace{-0.6mm} \xi \sum\nolimits_{c\in \mathcal{C}} x_{c,b}(t) I_c(t)A_{c,b}(t)\hspace{-0.1mm}-\hspace{-0.2mm}\kappa E^{Total}_{b}(t)\hspace{-0.2mm}-\hspace{-0.2mm}C^{Conf}\mathcal{F}_b(t)$ can be defined as the utility of ES $b\in\mathcal{B}$ in each time frame $t$, where we set $E^{Total}_{b}(t)\hspace{-0.5mm}=\hspace{-0.5mm}E^{Dtr}_{b}(t) \hspace{-0.5mm}+\hspace{-0.5mm} E^{Back}_{b}(t) \hspace{-0.5mm}+\hspace{-0.5mm} E^{Cre}_b(t) \hspace{-0.5mm}+\hspace{-0.5mm} E^{Mtr}_{b}(t)$ for simplicity, and $\mathcal{F}_b(t)\hspace{-0.5mm}=\hspace{-0.5mm}\{y_{b,n}(t)\}_{\forall n} \hspace{-0.5mm}\oplus\hspace{-0.5mm} \{y_{b,n}(t\hspace{-0.1mm}-\hspace{-0.1mm}1)\}_{\forall n}$ indicates whether sensors associated with ES $b\in \mathcal{B}$ change from $t\hspace{-0.1mm}-\hspace{-0.1mm}1$ to $t$.

Note that the equivalency between solving $[\mathcal{P}_1]$ and simultaneously solving $[\mathcal{P}_2]$ and $[\mathcal{P}_3]$ lies in the fact that i) decision variables in $[\mathcal{P}_2]$ and $[\mathcal{P}_3]$ together are identical to those in $[\mathcal{P}_1]$; ii) since $C^{Conf}\mathcal{F}(t)$ is irrelevant to the decision  $\{x_{c,b}(t)\}_{\forall t}$ in $[\mathcal{P}_2]$, the optimal solution that maximizes $\lim_{T\rightarrow \infty} \frac{1}{T} \sum\nolimits_{t = 0}^{T} U_{CS}(t)$ in $[\mathcal{P}_2]$ also maximizes $\mathcal{U}_{Sys}$ in $[\mathcal{P}_1]$, which can be written as $\mathcal{U}_{Sys} = \lim_{T\rightarrow \infty} \frac{1}{T} \sum\nolimits_{t = 0}^{T} U_{CS}(t) - C^{Conf}\mathcal{F}(t)$; and iii) since $E^{I}_{CS}(t)$ in the objective function of $[\mathcal{P}_1]$ is irrelevant to the decision $\{y_{b,n}(t), z_{b,n,w}(t), T_b(t)\}_{\forall t}$ in $[\mathcal{P}_3]$, the optimal solution that maximizes $\lim_{T\rightarrow \infty} \frac{1}{T}\sum\nolimits_{t = 0}^{T} \sum\nolimits_{b \in \mathcal{B}} U_{b}(t)$ in $[\mathcal{P}_3]$ also maximizes $\mathcal{U}_{Sys}$ in $[\mathcal{P}_1]$, which can be written as $\mathcal{U}_{Sys}=\lim_{T\rightarrow \infty} \frac{1}{T}\sum\nolimits_{t = 0}^{T} \sum\nolimits_{b \in \mathcal{B}} U_{b}(t) - \kappa E^{I}_{CS}(t)$.

\begin{figure}[!t]
	\centering
	\includegraphics[width=0.85\columnwidth]{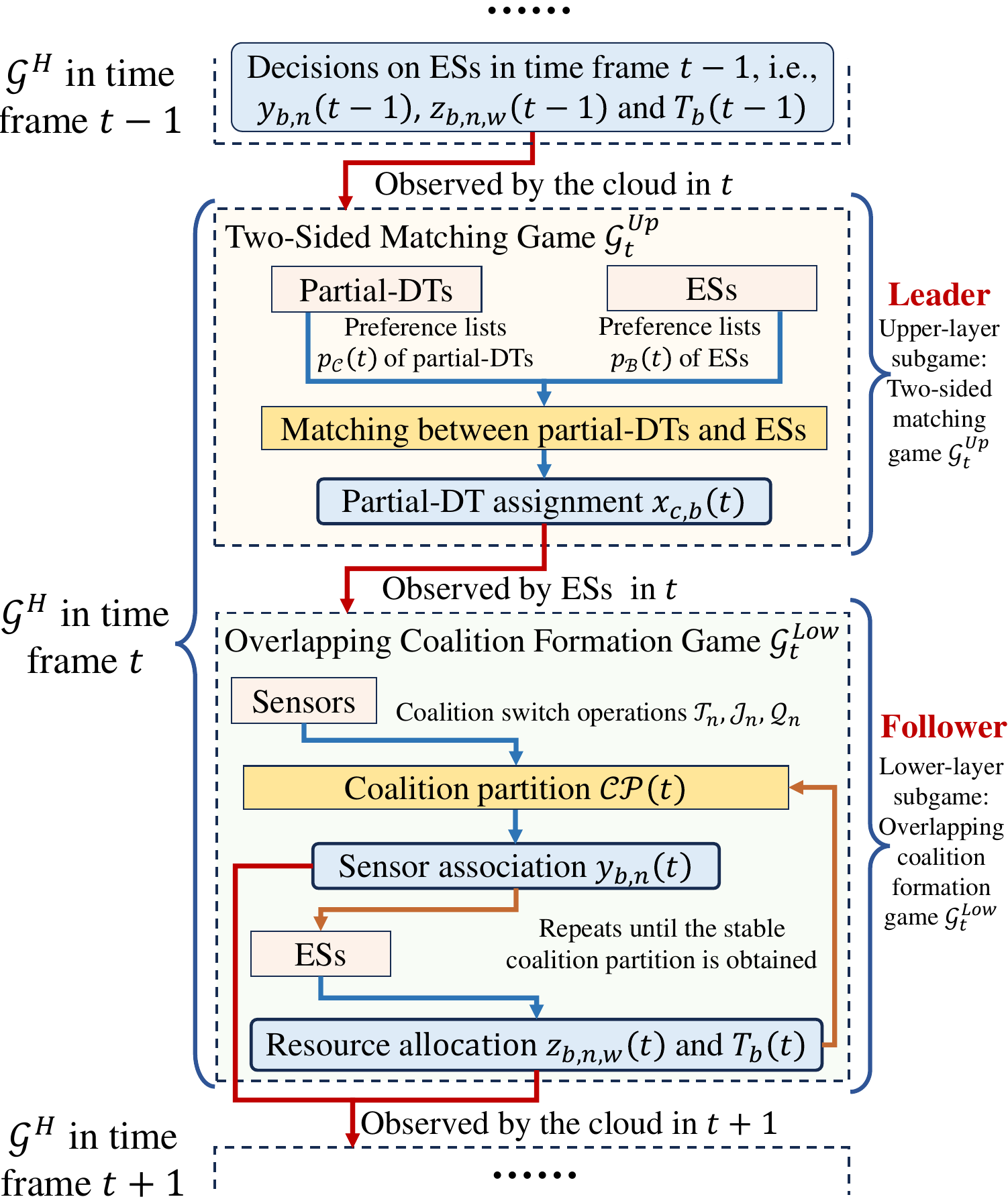}\\
	\vspace{-1.1mm}
	\caption{An illustration of the repeated and hierarchical decision-making process of $\mathcal{G}^H$. Within each time frame $t$, a two-sided matching game $\mathcal{G}^{Up}_t$ acts as the leader to determine partial-DT assignment $x_{c,b}(t)$, and based on this, an overlapping coalition formation game $\mathcal{G}^{Low}_t$ acts as the follower to determine ES-sensor association $y_{b,n}(t)$ and resource allocation decisions $z_{b,n,w}(t)$ and $T_b(t)$.
	}
	\label{gameformulation}
	\vspace{-1.5em}
\end{figure}

With the above transformation, we now aim to solve problems $[\mathcal{P}_2]$ and $[\mathcal{P}_3]$. To this end, we consider that the cloud acts as a dummy agent for solving $[\mathcal{P}_2]$ in strategically determining the optimal partial-DT assignments, while each ES acts as a dummy agent for solving $[\mathcal{P}_3]$ in strategically determining the optimal ES-sensor associations along with the optimal computation and communication resource allocations for partial-DT creations. Although these two decision-making processes are tightly coupled, it is clear that when ESs solve the ES-sensor associations and resource allocations, the partial-DT assignments have to be given in advance by the cloud, meaning that ESs can only make decisions after the cloud. This establishes a natural leader-follower relationship between the cloud and ESs in each time frame $t$. 
Therefore, as shown in Fig. \ref{gameformulation}, we formulate a two-layer hierarchical game $\mathcal{G}^H$ to solve $[\mathcal{P}_2]$ and $[\mathcal{P}_3]$ over the long run, where i) in the upper layer, a two-sided matching game $\mathcal{G}^{Up}_{t}$ is formulated to describe the cloud's problem $[\mathcal{P}_2]$ in each time frame $t$; and ii) in the lower layer, an overlapping coalition formation game $\mathcal{G}^{Low}_t$ is formulated to describe the ESs' problems $[\mathcal{P}_3]$ in each time frame $t$.
Unlike traditional hierarchical games in the existing literature \cite{OCFG1, BG2}, where all players' strategies are static, we let $\mathcal{G}^H$ evolve frame-by-frame, enabling the dynamic strategy adjustments. 
Specifically, in each time frame $t$, $\mathcal{G}^{Up}_t$ acts as the leader to determine the short-term decision $x_{c,b}(t)$, and based on this, $\mathcal{G}^{Low}_t$ acts as a follower to determine the short-term decisions $y_{b,n}(t)$, $z_{b,n,w}(t)$, and $T_b(t)$. 
Note that such a process repeats from one frame to another (as will be explained in subsection 4.3), integrating these two short-term subgames into a long-term hierarchical game framework of $\mathcal{G}^H$.

\begin{definition}[Two-layer hierarchical game $\mathcal{G}^H$ integrated with $\mathcal{G}^{Up}_{t}$ and $\mathcal{G}^{Low}_t$]
	Formally, $\mathcal{G}^H$ can be defined as
	\begin{equation}
		\begin{aligned}
			 \mathcal G^{H} = \big \langle G^H, \{\mathcal{G}^{Up}_{t}, \Pi^{Up}_t\}_{\forall t}, \{\mathcal{G}^{Low}_t, \Pi^{Low}_{t}\}_{\forall t}, \mathcal{U}_{CS},\mathcal{U}_{\mathcal{B}} \big \rangle, \label{GHdef1}
		\end{aligned}
	\end{equation}
	where $G^H = CS \cup \mathcal{B}$ represents the set of participants in $\mathcal{G}^H$, i.e., the cloud $CS$ and all ESs $\mathcal{B}$;
	$\mathcal{G}^{Up}_{t}$ stands for the upper-layer subgame in each time frame $t$, with $\Pi^{Up}_t$ being the corresponding strategy, and $\mathcal{G}^{Low}_t$ is the lower-layer subgame in each time frame $t$, with $\Pi^{Low}_{t}$ being the corresponding strategy; $\mathcal{U}_{CS}=\{U_{CS}(t)\}_{\forall t}$ and $\mathcal{U}_{\mathcal{B}}=\{U_{b}(t)\}_{\forall b,t}$ include the utilities of the cloud and ESs in all time frames, respectively.
\end{definition}

\begin{definition}[Upper-layer subgame: two-sided matching game $\mathcal{G}^{Up}_{t}$] For solving $[\mathcal{P}_2]$ in each time frame $t$, the decision on partial-DT assignments $x_{c,b}(t), \forall c \in \mathcal{C}, \forall b \in \mathcal{B}$, can actually be viewed as a bidirectional matching between partial-DTs in $\mathcal{C}$ and ESs in $\mathcal{B}$, such that each partial-DT has different preferences for being created on different ESs and vice versa \cite{GS0}. Hence, $\mathcal{G}^{Up}_{t}$ can be defined as
\begin{equation}
	\begin{aligned}
		\hspace{-1.5mm} \mathcal{G}^{U\hspace{-0.2mm}p}_{t} \hspace{-0.5mm} = \hspace{-0.5mm} \big \langle \hspace{-0.5mm} G^{U\hspace{-0.2mm}p}\hspace{-0.5mm}, \Pi^{U\hspace{-0.2mm}p}_t\hspace{-0.5mm}=\hspace{-0.5mm}\{\varPhi_t(i)\}_{\forall i\in G^{U\hspace{-0.2mm}p}}, \textbf{p}, U_{CS}(t),\{U_{b}(t)\}_{\forall b}\big \rangle,\label{GMdef}
	\end{aligned}
\end{equation}
where $G^{U\hspace{-0.2mm}p} = \mathcal{C} \cup \mathcal{B}$ denotes all partial-DTs and ESs involved in $\mathcal{G}^{U\hspace{-0.2mm}p}_{t}$; $\textbf{p}=\{\textbf{p}_c(t)=\{\{p^{c,b}\}_{\forall b}\},\textbf{p}_b(t)=\{p^{b,c}\}_{\forall b}\}_{\forall c,b}$ are the sets of all partial-DTs' and ESs' preference lists in time frame $t$, respectively, with $p^{c,b}, p^{b,c}\in [0,1]$ being the ranking value; $\varPhi_t(i), \forall i\in G^{U\hspace{-0.2mm}p}$ is the bidirectional matching between $\mathcal{C}$ and $\mathcal{B}$ in each time frame $t$, i.e.,
\begin{equation}
	\begin{aligned}
		\varPhi_t(c) = &~ b \Leftrightarrow \varPhi_t(b) = c, \text{if and only if } x_{c,b}(t) = 1, \\
		&\forall c \in \mathcal{C}, \forall b \in \mathcal{B}.
	\end{aligned}
\end{equation}
\end{definition}

\begin{definition}[Lower-layer subgame: overlapping coalition formation game $\mathcal{G}^{Low}_t$] For solving $[\mathcal{P}_3]$ in each time frame $t$, the decisions on ES-sensor associations $y_{b,n}(t)$ and resource allocations $z_{b,n,w}(t), T_b(t)$ can actually be viewed as an overlapping coalition formation among sensors and ESs. Specifically, all sensors are grouped into multiple overlapping coalitions, each of which associated with one ES, and each ES determines $z_{b,n,w}(t), T_b(t)$ based on the state of all coalitions. Sensors may strategically adjust their coalitions based on $z_{b,n,w}(t), T_b(t)$, and this process is conducted in an iterative manner till the convergence.
Hence, we define $\mathcal{G}^{Low}_t$ as
\begin{equation}
	\begin{aligned}
		\mathcal G^{Low}_{t}=&\big \langle G^{Low},\Delta, \Pi^{Low}_{t}  = \big\{\mathcal{CP}(t), \{\pi^R_b(t)\}_{\forall b}\big\},\\
		& \{U_{b}^{t}(Co_b)\}_{\forall b}, \{U_{n}(t)\}_{\forall n} \big \rangle,\label{GOdef}
	\end{aligned}
\end{equation}
where $G^{Low} = \mathcal{N} \cup \mathcal{B}$ denotes all sensors and ESs involved in $\mathcal{G}^{Low}_t$; $\Delta$ stands for the set of possible overlapping coalitions;  $\mathcal{CP}(t)=\{Co_b(t)\}_{\forall b \in \mathcal{B}}$ represents the coalition partition in time frame $t$, with $Co_b(t) = \{n \big| y_{b,n}(t)=1, \forall n \in \mathcal{N} \}$ being the coalition for ES $b \in \mathcal{B}$; $\pi^R_b(t) = \{z_{b,n,w}(t),T_b(t)\}_{\forall n,w}$ is the resource allocation strategy of ES $b\in\mathcal{B}$ in time frame $t$; 
$U_{b}^{t}(Co_b) = U_b(t), \text{if } Co_b = Co_b(t)$, shows the utility of ES $b\in\mathcal{B}$ in time frame $t$ given that its associated sensor coalition is $Co_b$, $U_{n}(t)=\sum\nolimits_{b \in \mathcal{B}}y_{b,n}(t)U^{Co_b}_{n}(t)$ can be seen as the coalitional utility of sensor $n\in \mathcal{N}$ in time frame $t$, with $U^{Co_b}_{n}(t)$ being the contribution of sensor $n$ to coalition $Co_b(t)$, calculated as the difference between the utility of ES $b$ with and without sensor $n$ in its associated coalition $Co_b$, i.e.,
\begin{align}
	\hspace{-0.8mm}&U^{Co_b}_{n}(t) = U^t_b(Co_b) - U^t_b(Co_b/n)  \notag\\
	\hspace{-0.8mm}&=x_{c,b}(t)\xi I_c(t) 
	\big [\varGamma \big (\hspace{-0.5mm}\sum\nolimits_{i=1}^t \hspace{-0.8mm}d_c(i) \hspace{-0.5mm} \big )\hspace{-0.5mm} \big (\hspace{-0.5mm} 1 \hspace{-0.5mm}-\hspace{-0.5mm} 2^{\left( T_b(t)(L\delta-2)\delta \gamma \right)/2} \big )  \notag\\
	\hspace{-0.8mm}&~~~~\hspace{-0.5mm}-\hspace{-0.5mm}\varGamma \big (\hspace{-0.5mm}\sum\nolimits_{i=1}^t d_c(i) \hspace{-0.5mm}-\hspace{-0.5mm} d_{n,c}(t) \hspace{-0.5mm}\big )\hspace{-0.5mm} \big (\hspace{-0.5mm} 1 \hspace{-0.5mm}-\hspace{-0.5mm} 2^{\frac{\hspace{-0.1mm}T_b\hspace{-0.1mm}(t)\hspace{-0.1mm}T^*_{b\hspace{-0.1mm}/\hspace{-0.1mm}n}\hspace{-0.1mm}(t)(\hspace{-0.1mm}L\delta \hspace{-0.1mm}-\hspace{-0.1mm}2)\delta \gamma}{2T^*_{b}(t)} }\hspace{-0.5mm} \big ) \big ]  \notag\\
	\hspace{-0.8mm}&~~~~\hspace{-0.5mm}-\hspace{-0.5mm}\kappa \big[\rho_b(F_b)^2\Upsilon_b \big (T_b(t)-{T_b(t)T^*_{b/n}(t)}/{T^*_{b}(t)} \big )  \notag\\
	\hspace{-1.0mm}&~~~~+ P_n x_{c,b}(t) d_{n,c}(t) \big / {R^T_{n,b}(t)}\big] \hspace{-0.5mm}-\hspace{-0.5mm} {C^{Conf}\mathcal{F}_b(t)}/{|\mathcal{B}|}. \label{UCobnt}
\end{align}
\end{definition}

\subsection{Game Analysis}
Let $(\Pi^{Up*}_t, \Pi^{Low*}_{t})_{\forall t}$ be the best response of the cloud and ESs in $\mathcal{G}^H$, i.e., the optimal strategies in solving $[\mathcal{P}_2]$ and $[\mathcal{P}_3]$ in every time frame $t$, then the equilibrium of $\mathcal{G}^H$ can be defined as follows.
\begin{definition}[Equilibrium of hierarchical game $\mathcal{G}^H$]
In $\mathcal{G}^H$, a strategy profile $(\Pi^{Up*}_t, \Pi^{Low*}_{t}\big )_{\forall t}$ is the equilibrium if $\Pi^{Up*}_t$ and $\Pi^{Low*}_{t}$ are respectively the equilibria of the upper-layer and lower-layer games in each time frame $t$, and moreover
\begin{equation}
	\begin{aligned}
		&\frac{1}{T} \sum\nolimits_{t=1}^{T} U_{CS}(t)|\Pi^{Up*}_t, \Pi^{Low*}_{t} \geq   \\
		&\frac{1}{T} \sum\nolimits_{t=1}^{T} U_{CS}(t)|\Pi^{Up}_t, \Pi^{Low*}_{t}, \text{ and}\\
		&\frac{1}{T} \sum\nolimits_{t=1}^{T} U_{b}(t)| \Pi^{Low*}_{t}, \Pi^{Up*}_t \geq   \\
		&\frac{1}{T} \sum\nolimits_{t=1}^{T} U_{b}(t)| \Pi^{Low}_{t}, \Pi^{Up*}_t, ~\forall b \in \mathcal{B}. \label{GHE0}
	\end{aligned}
\end{equation}
Clearly, when such an equilibrium is reached, the long-term utilities of both the cloud and ESs can be maximized so that $[\mathcal{P}_2]$ and $[\mathcal{P}_3]$ are solved, and no one will deviate this equilibrium unilaterally as it also guarantees the optimality in every time frame.
\end{definition}

\begin{definition}[Equilibrium of the upper-layer subgame $\mathcal{G}^{Up}_{t}$]\label{Stable_Matching}
	Given the equilibrium $\Pi^{Low*}_{t}$ of lower-layer subgame $\mathcal{G}^{Low}_{t}$, in $\mathcal{G}^{Up}_t$, the equilibrium $\Pi^{Up*}_t$ is defined as a stable matching $\{\varPhi^{*}_t(i)\}_{\forall i\in G^{U\hspace{-0.2mm}p}}$, which satisfies the condition that no blocking pair $(c, b)$ exists for any partial-DT $c \in \mathcal{C}$ and ES $b \in \mathcal{B}$. Particularly, a pair $(c, b)$ is blocking if both conditions $p^{c,b} > p^{c,\varPhi_t(c)}$ and $p^{b,c} > p^{b,\varPhi_t(b)}$ hold simultaneously, where $\varPhi_t(c)$ and $\varPhi_t(b)$ represent the current matched partners of $c$ and $b$ in time frame $t$, respectively. \end{definition}

\begin{definition}[Equilibrium of the lower-layer subgame $\mathcal G^{Low}_{t}$] \label{SOCP}
	In $\mathcal G^{Low}_{t}$, the equilibrium $\Pi^{Low*}_{t}$ is defined as the best response $\big (\mathcal{CP}^*(t), \{\pi^{R*}_b\}_{\forall b}\big )$, satisfying
	\begin{equation}
		\begin{aligned}
			&U_{b}(t)|\mathcal{CP}^*(t), \{\pi^{R*}_b(t)\}_{\forall b} \geq   \\
			&U_{b}(t)|\mathcal{CP}(t), \{\pi^{R*}_b(t)\}_{\forall b}, \text{ and}\\
			&U_{b}(t)|\{\pi^{R*}_b(t)\}_{\forall b}, \mathcal{CP}^*(t) \geq   \\
			&U_{b}(t)|\pi^{R}_b(t),\{\pi^{R*}_{b'}(t)\}_{\forall b' \neq b}, \mathcal{CP}^*(t), ~\forall b \in \mathcal{B}, \label{GLowE}
		\end{aligned}
	\end{equation}
	where $\mathcal{CP}^*(t)=\{Co^*_b(t)\}_{\forall b \in \mathcal{B}}$ represents the stable overlapping coalition partition of all sensors in time frame $t$. Here, we say the overlapping coalition partition $\mathcal{CP}^*(t)$ is stable if and only if the total utility of all ESs cannot be improved by unilaterally altering the coalition of any sensor even in an overlapped manner.
\end{definition}


%


Definitions 4-6 reveal an interdependence among $\mathcal{G}^H$, $\mathcal{G}^{Up}_{t}$, and $\mathcal{G}^{Low}_{t}$, such that the equilibrium of $\mathcal{G}^H$ depends on the equilibria of both $\mathcal{G}^{Up}_t$ and $\mathcal{G}^{Low}_t$ in each time frame $t$, while also extends them to accommodate to dynamic settings. Therefore, to facilitate the analysis, we first prove the existence of the equilibria in $\mathcal{G}^{Up}_t$ and $\mathcal{G}^{Low}_t$, and in turn prove the existence of the equilibrium in $\mathcal{G}^H$.


\begin{lemma}[Existence of the equilibrium in upper-layer subgame $\mathcal{G}^{Up}_{t}$]
Suppose the equilibrium $\Pi^{Low*}_{t}$ of lower-layer subgame $\mathcal{G}^{Low}_{t}$ exists, $\mathcal{G}^{Up}_{t}$ has at least one equilibrium $\Pi^{Up*}_t$, i.e., the stable matching $\{\varPhi_t(i)\}_{\forall i\in G^{U\hspace{-0.2mm}p}}$ in any time frame $t$.
\end{lemma}
\begin{IEEEproof}
	Please see Appendix A.
\end{IEEEproof}



\begin{lemma}[Existence of the equilibrium in lower-layer subgame $\mathcal G^{Low}_{t}$]
	$\mathcal G^{Low}_{t}$ has at least one equilibrium $\Pi^{Low*}_{t}$, i.e., a stable coalition partition $\mathcal{CP}^*(t)$ followed by optimal resource allocations $\{\pi^{R*}_b\}_{\forall b}$ in any time frame $t$.
\end{lemma}
\begin{IEEEproof}
	Please see Appendix B.
\end{IEEEproof}

\begin{theorem}[Existence of the equilibrium in hierarchical game $\mathcal{G}^H$]
	In $\mathcal{G}^H$, there always exists an equilibrium $(\Pi^{U\hspace{-0.2mm}p*}_t, \Pi^{Low*}_{t})_{\forall t}$ in long runs.
\end{theorem}
\begin{IEEEproof}
	Based on Lemmas 1 and 2, we aim to prove that $\mathcal{G}^H$ is a multi-player, general-sum, discounted-reward stochastic game, which naturally has a Markov perfect equilibrium (MPE) \cite{MPE0, MPE1}.
	
	First, we show that the strategy space for both the cloud and each ES is finite across multiple time frames in $\mathcal{G}^H$. 
	In terms of the cloud, the maximum cardinality of $\{\varPhi^{*}_t(i)\}_{\forall i\in G^{U\hspace{-0.2mm}p}}$ is $|\mathcal{C}|! \cdot {|\mathcal{B}|\choose |\mathcal{C}|}$. Meanwhile, in terms of each ES, the maximum cardinality of $z_{b,n,w}(t)$ is $|\mathcal{N}||\mathcal{W}|$, the maximum cardinality of $T_{b}(t)$ can be calculated as $\left \lceil \frac{2 }{(L\delta-2)\delta \gamma} \log_2(1 - \frac{A_c^{Req}}{ \varGamma \big (\sum\nolimits_{i=1}^t \min_{n,c} d_{n,c}(t) \big)}) \right \rceil$,
	and the maximum cardinality of $\mathcal{CP}^*(t)$ can be denoted as $2^{|\mathcal{B}||\mathcal{N}|}$. These indicate that the strategy spaces in $\mathcal{G}^H$ are all finite across multiple time frames.
	

	Furthermore, since strategies of the cloud and ESs in each time frame depend on themselves only and the resulted system states, e.g., the coalition structure and model accuracy, the state transitions in $\mathcal{G}^H$ satisfy the Markov property. Besides, it is clear that $\mathcal{G}^H$ is not a zero-sum game, as the total utility of the cloud and ESs does not sum to zero.
	Then, according to \cite{MPE0, MPE1}, we can conclude that $\mathcal{G}^H$ with finite strategy spaces is a multi-player, general-sum stochastic game. 
	
	In addition, referring to \cite{Discount1, myTWC}, maximizing the average infinite time-horizon utility and discounted infinite time-horizon utility are equivalent, and thus we have
	\begin{align}  
		\frac{1}{T}\sum\nolimits_{t=1}^{T}U_{x}(t) \Leftrightarrow \sum\nolimits_{t = 0}^{T} \eta^t U_{x}(t), \forall x \in \{CS \cup \mathcal{B}\}, \notag
	\end{align}
	where $\eta$ stands for the discount factor to balance the instant and future utilities. Given this fact, we can further conclude that $\mathcal{G}^H$ is a multi-player, general-sum, discounted-reward stochastic game in long runs, which must have an MPE. 
	This completes the proof.
\end{IEEEproof}

\subsection{A DRL-Based Solution Integrated with Stable Matching and Overlapping Coalition Formation (DMO)}
\begin{figure}[!t]
	\centering
	\includegraphics[width=0.9\columnwidth]{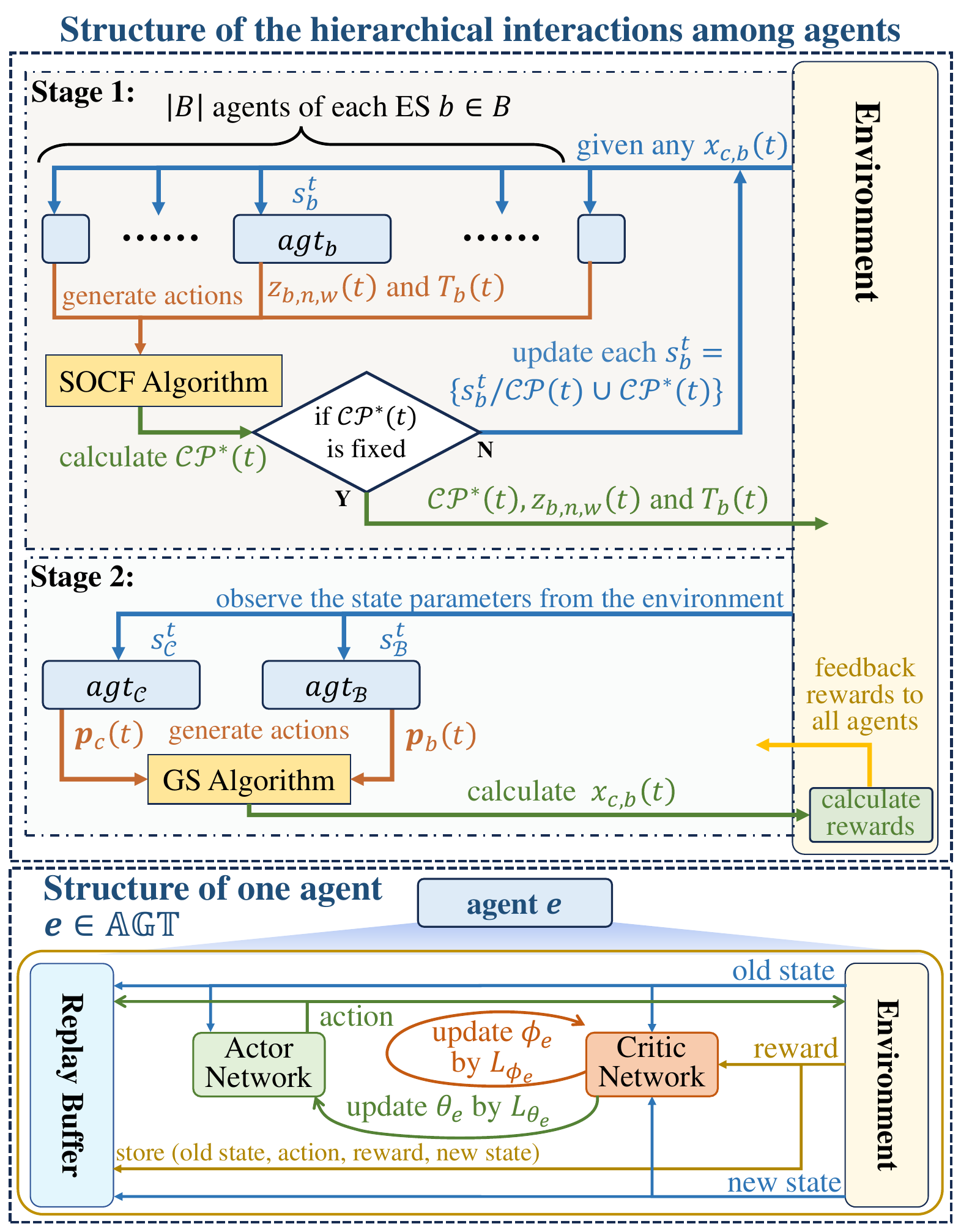}\\
	\vspace{-0.3mm}
	\caption{Overview of the proposed DMO approach.}\label{DRLStructure0}
	\vspace{-1.5em}
\end{figure}
In this subsection, we develop an SOCF algorithm and leverage the GS algorithm to find the equilibria of lower-layer subgame $\mathcal{G}^{Low}_t$ and upper-layer subgame $\mathcal{G}^{Up}_t$ within each single time frame $t$, respectively. 
We then propose a DRL-based solution to extend these two short-term equilibria into the long-term equilibrium of hierarchical game $\mathcal{G}^H$.  We name such a solution approach as DMO, with its overall structure illustrated by Fig. \ref{DRLStructure0}. Specifically, the proposed DMO operates in two stages.
In Stage 1, given any partial-DT assignments $x_{c,b}(t)$, the initial resource allocations $z_{b,n,w}(t)$ and $T_b(t)$ are derived by the actor networks of proximal policy optimization (PPO)-based agents in the DRL framework and then inputted into SOCF algorithm to obtain a temporary stable coalition partition $\mathcal{CP}^*(t)$. Iteratively, $\mathcal{CP}^*(t)$ is fed back into the PPO-based agents to update $\{\pi^*_b(t)\}_{\forall b}$ until the equilibrium of $\mathcal{G}^{Low}_t$, i.e., $(\mathcal{CP}^*(t), \{\pi^*_b(t)\}_{\forall b})$, converges.
In Stage 2, GS algorithm is employed to acquire the stable matching $\{\varPhi^{*}_t(i)\}_{\forall i\in G^{Up}}$, or equivalently the optimal partial-DT assignments $x^*_{c,b}(t)$, i.e., the equilibrium of $\mathcal{G}^{Up}_t$. This is achieved by inputting the preference lists $\{\textbf{p}_c(t)\}_{\forall c}$ and $\{\textbf{p}_b(t)\}_{\forall b}$, which are dynamically generated by the PPO-based agents in the DRL framework, informed by their observations of the system states. 
The whole process repeats frame-by-frame, allowing PPO-based agents to dynamically exploit historical knowledge and optimize strategies of both the cloud and ESs to achieve global optima in long runs, eventually reaching the equilibrium of $\mathcal {G}^H$ and thus solving the original problem.

%


%

\subsubsection{Short-Term Equilibrium Solutions to Subgames}



In each time frame $t$, the solution to $\mathcal{G}^{Low}_t$'s equilibrium $\big (\mathcal{CP}^*(t), \{\pi^{R*}_b\}_{\forall b}\big )$ should be first obtained due to the hierarchy that $\mathcal{G}^{Low}_t$ is the follower of $\mathcal{G}^{Up}_t$.
However, finding $\mathcal{CP}^*(t)$ is
much more difficult than traditional coalition games \cite{NonOverlap, 48}, because
in the considered system, sensors are allowed to form multiple overlapping coalitions, of which conditions for coalition partitions are hard to be evaluated.
To this end, we extend the conventional coalition switch rules \cite{NonOverlap} into overlapping coalition switch rules by particularly considering the condition that a sensor can join multiple coalitions as an overlap, and reformulate the utilities of overlapping coalition members. 
As such, we develop an SOCF algorithm to optimize $\mathcal{CP}^*(t)$ in one time frame $t$, producing the optimal $y_{b,n}(t)$, given the optimal $\{\pi^{R*}_b(t)\}_{\forall b}$ (derived by the actor network of PPO-based agents dynamically, which will be presented later). Note that the finally converged $\mathcal{CP}^*(t)$ and $\{\pi^{R*}_b(t)\}_{\forall b}$ together constitute $\mathcal{G}^{Low}_t$'s equilibrium introduced in Definition \ref{SOCP}.

\begin{definition}[Overlapping coalition switch rules]
	Let $Co_a, Co_b \in \Delta$ respectively denote sensor coalitions associated to any two ESs $a, b\in \mathcal{B}$ and $a\neq b$ in $\mathcal{G}^{Low}_t$, a sensor $n \in Co_a$ while $n \notin Co_b$ can be determined to change its coalition in an overlapped manner by i) transferring from $Co_a$ to $Co_b$, denoted as $\mathcal{T}_n(Co_a, Co_b)$, ii) joining $Co_b$ as an overlap, denoted as $\mathcal{J}_n(Co_b)$, and iii) unilaterally quitting $Co_a$, denoted as $\mathcal{Q}_n(Co_a)$, of which their respective conditions are as follows:
	
	\textbf{\textit{1) Transferring rule:}}
	$\mathcal{T}_n(Co_a, Co_b)$ is performed only when i) the remaining subcarriers of ES $b$ is sufficient, i.e., (\ref{PCB8}) holds, making it feasible; ii) the utility of $n$ can be increased, i.e., $U^{Co_b \cup n}_{n}(t) \geq \max\{0, U^{Co_a}_{n}(t) \}$; iii) the utility of all sensors except $n$ in $Co_b$ cannot be decreased, i.e., $U^{Co_b \cup n}_{\bar{n}}(t) \geq U^{Co_b}_{\bar{n}}(t), \forall \bar{n} \in Co_b$; and iv) the utility of all sensors in other coalitions that $n$ joins as an overlap cannot be decreased, i.e.,
	\begin{equation}
		\begin{aligned}
			&U^{Co_e}_{\bar{n}}(t) ~\Big |~ \mathcal{CP}(t)=\{Co_a/i, Co_b\cup n, Co_{-a,b}\} \geq  \\ 
			&U^{Co_e}_{\bar{n}}(t) ~\Big |~ \mathcal{CP}(t)=\{Co_a, Co_b, Co_{-a,b}\}, \\
			&\forall \bar{n} \in Co_e, \forall Co_e \in Co_{-a,b}, x_{e,n}=1, \label{TO4}
		\end{aligned}
	\end{equation}
	where $Co_{-a,b}$ stands for all coalitions in $\mathcal{CP}(t)$ other than $Co_a$ and $Co_b$.
	
	\textbf{\textit{2) Joining rule:}}
	$\mathcal{J}_n(Co_b)$ is performed only when i) resource constraints (\ref{sensorassociation}) and (\ref{PCB8}) are satisfied, making it feasible; ii) the utility of $n$ can be increased, i.e., $U^{Co_b \cup n}_{n}(t) \geq 0$; iii) the utility of all sensors except $n$ in $Co_b$ cannot be decreased, i.e., $U^{Co_b \cup n}_{\bar{n}}(t) \geq U^{Co_b}_{\bar{n}}(t), \forall \bar{n} \in Co_b$; and iv) the utility of all sensors in other coalitions that $n$ joins as an overlap cannot be decreased, i.e.,
	\begin{equation}
		\begin{aligned}
			&U^{Co_e}_{\bar{n}}(t) \Big | \mathcal{CP}(t)=\{Co_b\cup n, Co_{-b}\} \geq \\ 
			&U^{Co_e}_{\bar{n}}(t) \Big | \mathcal{CP}(t)=\{Co_b, Co_{-b}\}, \\
			&\forall \bar{n} \in Co_e, \forall Co_e \in Co_{-b}, x_{e,n}=1. \label{JO5}
		\end{aligned}
	\end{equation}
	
	\textbf{\textit{3) Quitting rule:}}
	$\mathcal{Q}_n(Co_a)$ is performed only when i) the utility of $n$ can be increased, i.e., $U^{Co_a}_{n}(t) \leq 0$; ii) the utility of all sensors except $n$ in $Co_a$ cannot be decreased, i.e., $U^{Co_a / n}_{\bar{n}}(t) \geq U^{Co_a}_{\bar{n}}(t), \forall \bar{n} \in Co_a$.
\end{definition}

\begin{algorithm}[t!]
	\footnotesize
	\caption{SOCF Algorithm for $\mathcal{CP}^*(t)$} 
	\KwIn{Partial-DT assignment $x_{c,b}(t)$, initial decisions on $z_{b,n,w}(t)$ and $T_b(t)$, previous decision on $y_{b,n}(t-1)$;}
	\KwOut{Stable coalition partition $\mathcal{CP}^*(t)$, or equivalently the optimal ES-sensor associations $y^*_{b,n}(t)$;}
	\For{ES $b \in \mathcal{B}$}{
		ES $b$ calculates $\mathbb{N}_b(t)$ according to constraint (\ref{ESaccess})\;
		ES $b$ selects all sensors in $Co_b(t-1)$ and randomly selects at most $\max\{|\mathcal{W}|- |Co_b(t-1)|\}$ sensors in $\mathbb{N}_b(t)$. 
	}
	All ESs generate the initial overlapping coalition partition $\mathcal{CP}(t)$\;
	\While{$\mathcal{CP}(t)$ is different from it in the previous loop}{
		\For{$n \in \mathcal{N}$}{
			\For{$Co_a \in \{Co_x \big | n \in Co_x \}$}{
				\For{$Co_b \notin \{Co_x \big | n \in Co_x \}$}{
					\If{\textit{Transferring rule} is satisfied}{
						$\mathcal{T}_n(Co_a, Co_b)$ is performed\;
					}
					\ElseIf{\textit{Joining rule} is satisfied}{
						$\mathcal{J}_n(Co_a, Co_b)$ is performed\;
					}
					\ElseIf{\textit{Quitting rule} is satisfied}{
						$\mathcal{Q}_n(Co_a)$ is performed\;
					}
				}
			}
			Update $\mathcal{CP}(t)$, all ESs synchronize such information\;
		}
	}
	According to final $\mathcal{CP}(t)$, all ESs update $y_{b,n}(t)$.
\end{algorithm}

We are now able to design an SOCF algorithm based on the above coalition switch rules to obtain the stable coalition partition $\mathcal{CP}^*(t)$ in lower-layer subgame $\mathcal{G}^{Low}_t$. Specifically, in each time frame $t$, given the optimal resource allocations $\{\pi^{R*}_b(t)\}_{\forall b}$ and partial-DT assignments $x_{c,b}(t)$, each ES initializes its ES-sensor association by i) calculating the set of eligible sensors $\mathbb{N}_b(t)=\{n ~\big |~ R^T_{n,b}(t)\geq R^T_{Min}\}$, where each sensor satisfies (\ref{ESaccess}); ii) selecting all sensors in its previous coalition $Co_b(t-1)$; and iii) randomly selecting at most $\max\{|\mathcal{W}|- |Co_b(t-1)|\}$ sensors in $\mathbb{N}_b(t)$ without violating (\ref{sensorassociation}) and (\ref{PCB8}). Then the initial $\mathcal{CP}(t)$ is formed, and $|\mathcal{N}|$ sensors are required to perform coalition switch operations.
After initialization, each sensor simultaneously performs switch operations, i.e., $\mathcal{T}_n(Co_a, Co_b), \mathcal{J}_n(Co_b)$ and  $\mathcal{Q}_n(Co_a)$, adhering to the switch rules defined in Definition 7. 
This process repeats until no sensor has the motivation to perform any further switch operations, i.e., the coalition partition converges to $\mathcal{CP}^*(t)$. Correspondingly, the optimal ES-sensor associations, i.e., $y^*_{b,n}(t), \forall b \in \mathcal{B}, \forall n \in \mathcal{N}$, can be obtained from $\mathcal{CP}^*(t)$. The detailed procedure of this algorithm is summarized in Algorithm 1.

\begin{theorem}[Convergence and optimality of SOCF]
	 Starting from any initial overlapping coalition partition, the proposed SOCF algorithm can eventually converge to a stable overlapping coalition partition $\mathcal{CP}^*(t)$ in each time frame $t$, yielding the optimal ES-sensor associations.
\end{theorem}
\begin{IEEEproof}
	Please see Appendix C.
\end{IEEEproof}

Then, given the equilibrium solution of $\mathcal{G}^{Low}_t$, for the upper-layer subgame $\mathcal{G}^{Up}_t$, since it is a standard one-to-one matching game, we employ the well-known GS algorithm \cite{GS0, GS1} to achieve the equilibrium, i.e., the stable matching $\{\varPhi^*_t(i)\}_{\forall i \in G^{Up}}$ in each time frame $t$, of which the input is preference lists $\{\textbf{p}_c(t)\}_{\forall c}$ and $\{\textbf{p}_b(t)\}_{\forall b}$. 
As GS algorithm has been widely proved with the guarantee of convergence to a stable matching, $\{\varPhi^*_t(i)\}_{\forall i \in G^{Up}}$ can certainly be obtained, leading to the equilibrium of $\mathcal{G}^{Up}_t$.


\subsubsection{Long-Term Equilibrium Solution to Hierarchical Game}
Having the short-term equilibria of subgames $\mathcal{G}^{Up}_t$ and $\mathcal{G}^{Low}_t$ in each time frame $t$, we can extend them into the long-term equilibrium of hierarchical game $\mathcal{G}^H$. Note that the long-term decision-makings between any two consecutive time frames are inherently coupled, particularly, i) the decisions on preference orders $\textbf{p}_c(t), \textbf{p}_b(t)$ affect the partial-DT assignments $x_{c,b}(t)$ determined by the cloud, which subsequently affects the coalition partition $\mathcal{CP}(t)$ and resource allocation $\pi_b^R(t)$ determined by each ES $b \in \mathcal{B}$; and ii) the decisions on resource allocations $\{\pi_b^R(t)\}_{\forall b}$ impacts $\mathcal{CP}(t)$ in each time frame $t$, which in turn affects $\textbf{p}_c(t+1), \textbf{p}_b(t+1)$ in the next time frame $t+1$. To tackle such an issue, we first model the long-term decision-makings on resource allocations and preference lists as two coupled Markov decision processes (MDPs), and then develop a DMO approach to dynamically solve these MDPs. Particularly, given the outputs of PPO-based agents, we call SOCF and GS algorithms to derive $\mathcal{CP}^*(t)$ and $\{\varPhi^*_t(i)\}_{\forall i \in G^{Up}}$ in each time frame $t$, and then we continuously train these agents using the historical knowledge, dynamically updating their outputs. The final outputs of DMO, including $\{\pi_b^R(t)\}_{\forall b}$ and $\textbf{p}_c(t), \textbf{p}_b(t)$ from PPO-based agents, and $\mathcal{CP}(t)$ and $\{\varPhi^*_t(i)\}_{\forall i \in G^{Up}}$ from the SOCF and GS algorithms within the DRL framework, together constitute the long-term equilibrium of $\mathcal{G}^H$, or equivalently the optimal solution to the original problem $[\mathcal{P}_1]$.

In the following, we formulate the MDPs for the decision-makings on resource allocations $\{\pi_b^R(t)\}_{\forall b}$ and preference lists $\textbf{p}_c(t), \textbf{p}_b(t)$ as $\mathcal M_{R}=\{\mathcal{S}_{R},\mathcal{A}_{R},\Xi_{R},\mathcal R_{R}\}$ and $\mathcal M_{PL}=\{\mathcal{S}_{PL},\mathcal{A}_{PL},\Xi_{PL},\mathcal R_{PL}\}$, respectively, in detail.

\textit{1) Environment States}: 
In each time frame $t$, the environment states in MDP $\mathcal M_{R}$ is the observation on system dynamics in the view of each ES $b \in \mathcal{B}$ in $\mathcal{G}^{Low}_t$, i.e., $s_{b}^t=\big\{\{I_c(t), d_c(t\hspace{-0.3mm}-\hspace{-0.3mm}1), \sum_{i=1}^{t-1}d_c(i)\}_{\forall c}, \{\varPhi^*_t(i)\}_{\forall i \in G^{Up}},\mathcal{CP}(t)\big\}$.
Meanwhile, the environment states in MDP $\mathcal M_{PL}$ is the observations on system dynamics in the view of all partial-DTs $\mathcal{C}$ and all ESs $\mathcal{B}$ in $\mathcal{G}^{Up}_t$, which can be defined as $s_{\mathcal{C}}^t=\big\{\{I_c(t), d_c(t\hspace{-0.4mm}-\hspace{-0.4mm}1), \sum_{i=1}^{t-1}d_c(i) \}_{\forall c}, \{\pi_b^R(t\hspace{-0.4mm}-\hspace{-0.4mm}1)\}_{\forall b}, \mathcal{CP}(t\hspace{-0.4mm}-\hspace{-0.4mm}1) \big\}$,
and $s_{\mathcal{B}}^t=\big\{\{I_c(t), d_c(t\hspace{-0.3mm}-\hspace{-0.3mm}1),\sum_{i=1}^{t-1}d_c(i)\}_{\forall c}, \{\varPhi^*_t(i)\}_{\forall i \in G^{Up}}\big\}$, respectively. For conciseness, let $\mathcal{S}_{PL} = \{s_{\mathcal{C}}^t, s_{\mathcal{B}}^t\}_{\forall t}$ and $\mathcal{S}_{R} = \{s_{b}^t\}_{\forall b,t}$ denote the state space of MDPs $\mathcal M_{R}$ and $\mathcal M_{PL}$, respectively.

\textit{2) Actions}: In each time frame $t$, the actions in MDPs $\mathcal M_{R}$ and $\mathcal M_{PL}$ are exactly the resource allocations and preference lists, and thus we have their respective action spaces $\mathcal{A}_{R}=\{a^t_b\}_{\forall b,t}=\{z_{b,n,w}(t), T_b(t)\}_{\forall b,t}$ and $\mathcal{A}_{PL}=\{a^t_{\mathcal{C}}, a^t_{\mathcal{B}}\}_{\forall t}=\{ \{\textbf{p}_c(t)\}_{\forall c}, \{\textbf{p}_b(t)\}_{\forall b}\}_{\forall t}$.

\textit{3) State Transition Probabilities}: The state transition probability from state $s\in \mathcal{S}_i, \forall i \in \{R,PL\}$ to another state $s' \in \mathcal{S}_i$ by taking action $a_i\in \mathcal{A}_i$ is expressed as $\Xi_i^{s,s'}=Pr(s'|s,a_i) \in [0,1]$.
 
\textit{4) Rewards}: The immediate rewards for ES $b \in \mathcal{B}$ in $\mathcal M_{R}$ can be calculated as
\begin{equation}
	\begin{aligned} 
		r_{b}^t=
		\left\{
		\begin{array}{l}
			U_{b}(t), \text{if $\tau^{Total}(t) \leq \delta_T$}, \\
			~~~~~~0, \text{otherwise},
		\end{array}\right.
	\end{aligned} \label{MDPR}
\end{equation}
and that for $i,\forall i = \mathcal{C} \text{ or } \mathcal{B}$ in $\mathcal M_{PL}$ can be written as
\begin{equation}
	\begin{aligned} 
		r_{i}^t=
		\left\{
		\begin{array}{l}
			~~~~~~~U_{CS}(t), \text{if $i=\mathcal{C}$ and if $\tau^{Total}(t) \leq \delta_T$},\\
			\sum\nolimits_{b \in \mathcal{B}} U_{b}(t), \text{if $i= \mathcal{B}$ and if $\tau^{Total}(t) \leq \delta_T$},\\
			~~~~~~~~~~~~~~~~\hspace{0.5mm}0, \text{otherwise}.
		\end{array}\right.
	\end{aligned} \label{MDPR1}
\end{equation}

Based on the above formulation, we next elaborate the DMO approach in terms of its agent structure, training workflow, and network parameter updates.

First, the decisions in either $\mathcal M_{R}$ or $\mathcal M_{PL}$ are multi-dimensional, making it computationally intensive to explore the vast action space. To address this complexity, we assign the actions in $\mathcal A_{R}$ and $\mathcal A_{PL}$ to multiple agents, which enhances exploration efficiency and accelerates the training process.
Particularly, we introduce $|\mathcal{B}|$ agents $agt_b, \forall b \in \mathcal{B}$, where each agent $agt_b$ generates resource allocation $\pi^R_b(t)$. Additionally, agents $agt_{\mathcal{C}}$ and $agt_{\mathcal{B}}$ are responsible for generating $\{\textbf{p}_c(t)\}_{\forall c}$ and $\{\textbf{p}_{b}(t)\}_{\forall b}$, respectively.
The set of all agents is denoted as $\mathbb{AGT}$. To improve the training efficiency, we employ the actor-critic (AC) framework \cite{CJY0}.  
For each agent $e \in \mathbb{AGT}$, we design i) a critic network with parameter $\phi_e$ to estimate state-value $V_{\phi_e}(s)\hspace{-0.5mm}\approx \hspace{-0.5mm}V_{e}(s)$ of $e$, where $V_{e}(s)=\mathbb{E}\{\sum_{t=1}^T \eta^t r_e^t|s_0=s\}$ indicates the long-term discounted reward for $e$ with discount factor $\eta$ and initial state $s_0$; ii) an actor network with parameter $\theta_e$ to parameterize the optimal strategy of the corresponding agent; and iii) a replay buffer to store the historical knowledge, including the previous states, new states, actions and rewards, during the DRL training process.

Second, since the formulated MDPs are inherently coupled, we further propose a hierarchical training process in DRL, where the interaction among all agents adhere to the decision-making sequence in $\mathcal{G}^H$. 
Specifically, in each time frame $t$ within one training step, such interactions present two stages,
i) in stage 1, the agents $agt_{b}, \forall b \in \mathcal{B}$ simultaneously generate $\pi^R_b(t)$ based on $s^t_{b}$ with initial coalition partition $\mathcal{CP}(t-1)$, then we call the SOCF algorithm to iteratively obtain the stable coalition partition and update $s^t_{b}$ for each agent $agt_{b}$, until both $\mathcal{CP}(t)$ and $\pi^R_b(t)$ converge;
and ii) in stage 2, $agt_{\mathcal{C}}$ and $agt_{\mathcal{B}}$ first generate $\{\textbf{p}_c(t)\}_{\forall c}$ and $\{\textbf{p}_{b}(t)\}_{\forall b}$ based on $s^t_{\mathcal{C}}$ and $s^t_{\mathcal{B}}$, respectively, and then we call GS algorithm to obtain $\varPhi^*_t(c)$.
Note that original reward functions in (\ref{MDPR}), (\ref{MDPR1}) may introduce the problem of sparse rewards \cite{SR}, leading to a lack of sufficient exploration on both the action and state spaces. To handle this, we introduce a penalty for violating constraint (\ref{RealTime}) to the reward functions, i.e., $\max\big\{\psi \big (\tau^{Total}(t)-\delta_T \big ), 0\big\}$, where $\psi$ denotes the penalty coefficient.
Accordingly, we can calculate the rewards of all agents in time frame $t$, and then store the tuple $\{s^{t-1}_{e}, a^t_{e}, s^{t}_{e}, r^{t}_{e}\}$ in replay buffer for each agent $e \in \mathbb{AGT}$.


Finally, network parameters of both the actor and critic networks within each agent should be updated at a certain frequency for achieving the global optima. 
For each agent $e\in \mathbb{AGT}$, we employ the PPO method \cite{myTWC} in network updating, to fully explore the time-varying environment states and efficiently exploit the historical knowledge. Specifically, the PPO-based updating process includes i) calculating the rewards-to-go of $e$, i.e., the discounted rewards $J_e(t)=\sum^{T}_{t=1} \eta^{t}r_e^t$; ii) calculating $e$'s advantage function $A_e(t)=J_e(t)-V_{\phi_e}(s_e^t)$; iii) calculating the loss function of $e$'s actor network: $L_{\theta_e}\hspace{-0.5mm}=\hspace{-0.5mm}\sum_{t=1}^{T} A_e(t)\min ( \frac{p_{\theta_e}(a_e^t|s_e^t)}{p_{\theta'_e}(a_e^t|s_e^t)},clip(\frac{p_{\theta_e}(a_e^t|s_e^t)}{p_{\theta'_e}(a_e^t|s_e^t)},1-\epsilon,1+\epsilon))$, where $p_{\theta_e}(a_e^t|s_e^t)$ is the probability of the $e$'s actor network with parameter $\theta_e$ choosing action $a_e^t$ at state $s_e^t$, $\theta'_e$ is the parameter of the $e$'s actor network before updating, and $clip(\cdot)$ is the clip function with predefined hyperparameter $\epsilon$; and iv) calculating the loss function of the $e$'s critic network: $L_{\phi_e}\hspace{-1.5mm}=\hspace{-0.5mm}\sum_{t=1}^{T} (V_{\phi_e}(s_e^t)\hspace{-0.5mm}-\hspace{-0.5mm}J_e(t))^2$. Then, the network parameters $\theta_e$ and $\phi_e$ for each agent $e \in \mathbb{AGT}$ can be updated for minimizing their corresponding loss functions $L_{\theta_e}$ and $L_{\phi_e}$ via random gradient decedent methods, e.g. Adam optimizer \cite{AC0}.

In summary, all detailed steps of the proposed DMO approach are presented in Algorithm \ref{algorithm}.
\begin{theorem}[Computational complexity of DMO]
	The computational complexity of the proposed DMO approach is 
	$\mathcal{O}\big(\textbf{I} \big( 2T(|\mathcal{S}_{PL}|+|\mathcal{A}_{PL}|)+|\mathcal{B}|T(|\mathcal{S}_{R}|+|\mathcal{A}_{R}|) + 6K^2_{hid}  + |\mathcal{C}||\mathcal{B}| + \varpi |\mathcal{N}|L_n(|\mathcal{B}|-L_n) \big) \big)$, where $\textbf{I}$ is the maximum number of training steps, and $\varpi$ denotes the maximum number of times to call the SOCF algorithm within each training step.
\end{theorem}
\begin{IEEEproof}
	Please see Appendix D.
\end{IEEEproof}



\begin{algorithm}[t!]
	\footnotesize
	\caption{DMO Approach for $(\Pi^{Up*}_t, \Pi^{Low*}_{t}\big )_{\forall t}$} \label{algorithm}
	\KwIn{Locations of all devices, importance weights $\{I_c(t)\}_{\forall c}$.}
	\KwOut{The optimal strategies in $\mathcal{G}^H$, i.e., $(\Pi^{Up*}_t, \Pi^{Low*}_{t})$.}
	Randomly initialize: Parameters of both the actor and critic networks within all agents, all decision variables and $\mathcal{CP}(1)$\;
	\For{$training ~step = 1,2,..., max ~steps$ }{
		\For{ $t = 1,2,...,T$ }{
			$\mathcal{CP}(t)\leftarrow\mathcal{CP}^*(t-1)$\;
			Calculate $s_{b}^t, \forall b \in \mathcal{B}$\;
			\For{Given all possible condition of $\{x_{c,b}(t)\}_{\forall c,b}$}{
			\While{$\mathcal{CP}^*(t)$ is not fixed}{
				Each agent $agt_{b}, \forall b \in \mathcal{B}$ generates action $\pi^R_b(t)$ based on $s^t_b$, respectively\;
				Call SOCF Algorithm to get $\mathcal{CP}^*(t)$\;
				$\mathcal{CP}(t) \leftarrow \mathcal{CP}^*(t)$\;
			}}
			Calculate $s_{\mathcal{C}}^t$ and $s_{\mathcal{B}}^t$\;
			$agt_{CS}$ and $agt_{\mathcal{B}}$ generate actions $\{\textbf{p}_{c}(t)\}_{\forall c}$ and $\{\textbf{p}_{b}(t)\}_{\forall b}$ based on $s_{CS}^t$ and $s_{\mathcal{B}}^t$, respectively\;
			Call GS algorithm to calculate $\{x_{c,b}(t)\}_{\forall c,b}$\;
	
			Calculate the reward $r_e^t$ of each agent $e \in \mathbb{AGT}$\;
			Update the state $s_e^{t+1}$ of each agent $e \in \mathbb{AGT}$\;
			\For{each agent $e\in \mathbb{AGT}$}{
				Store tuple $\{s_e^t, a_e^t, r_e^t, s_e^{t+1}\}$ in $e$'s replay buffer\;
			}	
		}
		\If{$training ~step ~\%~ update ~steps = 0$}{
			\For{each agent $e\in \mathbb{AGT}$}{
				Calculate $e$'s rewards-to-go $J_e(t)=\sum^{T}_{t=1} \eta^{t}r_e^t$\;
				Calculate $e$'s advantage $A_e(t)=J_e(t)-V_{\phi_e}(s_e^t)$\;
				Generate the loss for $e$'s actor network: $L_{\theta_e}\hspace{-0.5mm}=\hspace{-0.5mm}\sum_{t=1}^{T} A_e(t)\min ( \frac{p_{\theta_e}(a_e^t|s_e^t)}{p_{\theta'_e}(a_e^t|s_e^t)},clip(\frac{p_{\theta_e}(a_e^t|s_e^t)}{p_{\theta'_e}(a_e^t|s_e^t)},1-\epsilon,1+\epsilon))$\;
				Generate the loss for $e$'s critic network: $L_{\phi_e}\hspace{-1.5mm}=\hspace{-0.5mm}\sum_{t=1}^{T} (V_{\phi_e}(s_e^t)\hspace{-0.5mm}-\hspace{-0.5mm}J_e(t))^2$\;
				Update $\theta_e$ and $\phi_e$ via Adam Optimizer.
			}
		}
	}
\vspace{-0.5em}
\end{algorithm}

\section{Simulation Results}\label{SR}
In this section, simulations are conducted to evaluate the performance of the proposed DMO approach in optimizing the long-term system performance for the federated DT construction. All simulation results are obtained by taking average over 1000 runs with various parameter settings.
\subsection{Simulation Settings}

\begin{table}[!t]
	\vspace{-0.7em}
	\footnotesize
	\caption{Simulation Parameters}
	\vspace{-1.0em}
	\renewcommand{\arraystretch}{1.1}
	\label{table2}
	\centering
	\resizebox{\columnwidth}{!}{
		\begin{tabular}{l|l|l|l}
			\hline
			\bfseries Parameter&\bfseries Value & \bfseries Parameter &\bfseries Value\\ 
			\hline	
			$I_c(t)$& $[0,1]$ & $L_n$ & $3$ \\
			$W$ & $[1, 5]$~MHz & $|\mathcal{W}|$ & $10$\\
			$\beta$ & $200$ & $\rho_b$ &  $10^{-16}$ \\
			$L,\delta,\gamma$ & $\{8, 0.02, 2\}$ & $F_b, F_{CS}$ & $\{64, 3000\}$~MHz \\
			$R^T_{CS,b}, R^T_{b,CS}, R^T_{b,b'}$ & $[1,3]$~Mbps & $\Upsilon_b$ & 120~cycles/bit \\
			$R^T_{Min}$ & $ [1,5]$~kbps & $d_{n,c}(t)$ & $[200, 600]$~kbits \\
			$D_c$ & $[1, 5]$~Mbits & $P_b,P_n$ & $[5, 33]$~dBm \\
			$P^{Max}_{CS}, P^{Idle}_{CS}, P^{Leak}_{CS}$ & $[1, 60]$~Watt & $n^2_b$ & $-104$~dbm/Hz \\
			$\xi,\kappa, C^{Conf}$ & $[0.1, 15]$ & $\delta_T$ & $7.6$s \\	
			$H_{n,b,w}(t)$ & $\{0.2, 0.4, 0.6\}$ & $\eta$& 0.92 \\		
			\hline
	\end{tabular}}
	\vspace{-0.5em}
\end{table}
\begin{table}[!t]
	\footnotesize
	\caption{Neural Network Settings}
	\vspace{-1.0em}
	\renewcommand{\arraystretch}{1.2}
	\label{table3}
	\centering
	\begin{tabular}{c|c|c}
		\hline
		\multicolumn{3}{c}{\bfseries Actor Network}\\
		\hline
		\multicolumn{3}{c}{Linear layers with size $\{(K_e^I, 64),(64,64),(64,K_e^O)\}$ }\\
		\hline
		\multicolumn{3}{c}{\bfseries Critic Network}\\
		\hline
		\multicolumn{3}{c}{Linear layers with size $\{(K_e^I, 64),(64,64),(64,1)\}$ }\\
		\hline
		\bfseries{Agent $e$} & \bfseries{Input Size $K_e^I$} & \bfseries{Output Size $K_e^O$}\\
		\hline
		$agt_{\mathcal{C}}$ & $3|\mathcal{C}| + |\mathcal{B}| \big(1 + |\mathcal{N}|(|\mathcal{W}|+ 1) \big) $ & $|\mathcal{B}||\mathcal{C}|$ \\
		$agt_{\mathcal{B}}$ & $|\mathcal{C}|(4+|\mathcal{B}|)$ & $|\mathcal{B}||\mathcal{C}|$ \\
		$agt_{b}$ & $(4+|\mathcal{B}|) + |\mathcal{B}||\mathcal{N}|$ & $|\mathcal{N}|+1$ \\
		\hline
	\end{tabular}
	\vspace{-1.2em}
\end{table}

We consider a federated DT construction framework enabled by \textcolor{black}{distributed} sensing within a $1000m\times1000m$ geographic area, where $|\mathcal{N}|=20$ sensors and $|\mathcal{B}|=5$ ESs are randomly scattered, aiming to build a global DT model consisting of $|\mathcal{C}| = 5$ partial-DTs. All key simulation parameters are listed in Table 2, most of which are set according to conventional configurations in the literature \cite{FL0, AC0, OCFG1, BG1, ssfTNSE, syTWC}. Besides, the structure of actor and critic networks, along with hyperparameter settings for the proposed DMO approach, are provided in Table \ref{table3}. Note that some parameters may vary for different evaluation purposes.

Furthermore, to show the effectiveness of the proposed federated DT construction framework, several alternative DT construction schemes are simulated for comparison.
\begin{itemize}
	\item $Centra$ \cite{Centra}:
	All sensing data is directly uploaded from sensors to the central cloud for constructing the global DT model.
	\item $Non\mbox{-}Overlap$ \cite{NonOverlap}:
	A similar edge-cloud collaborative federated DT construction framework is utilized while each sensor is associated with only one ES in each time frame, disabling potential overlaps.
\end{itemize}

\textcolor{black}{Additionally, to evaluate the superiority of the proposed DMO approach in optimizing the long-term performance for our considered federated DT construction framework, the following algorithms are simulated as benchmarks.
	\begin{itemize}
		\item $GRE$: All optimization decisions are greedily determined to maximize the utility function $\mathcal{U}(t)$ within each single time frame, ignoring the interdependence between different time frames.
		\item $QL$ \cite{Rsp39}: All optimization decisions are determined to maximize state-action values (i.e., Q-values) frame-by-frame, which are stored and updated by predefined tables rather than deep neural networks.
		\item $LBCD$ \cite{BGour2}: All optimization decisions are relaxed to continuous form and dynamically determined by block coordinate descent algorithm, by approximately decomposing the original problem by Lyapunov method into instant problem in each time frame.
		\item $PTS$ \cite{OCFG5}: All optimization decisions are determined using preference gravity-based tabu search algorithm within each single time frame, except that ES-sensor associations are optimized frame-by-frame. 
		\item $MAB$ \cite{FL0}: All optimization decisions are made dynamically by the contextual multi-armed bandit algorithm (MAB), where environment states serve as contexts to guide the long-term decision-makings by MAB-based agents.		
	\end{itemize}
}

\subsection{Performance Evaluations}

\begin{figure}[!t]
	\centering
	\includegraphics[width=0.8\linewidth]{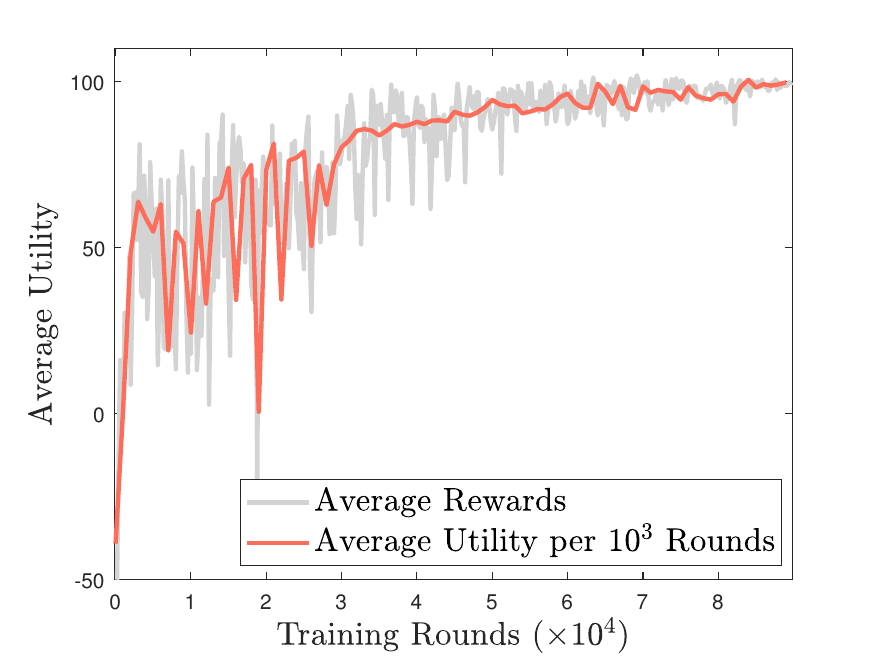}\\
	\vspace{-0.3em}
	\caption{Convergence of the proposed DMO approach.} \label{simu0}
	\vspace{-1.2em}
\end{figure}
\begin{figure}[!t]
	\centering
	\includegraphics[width=0.8\linewidth]{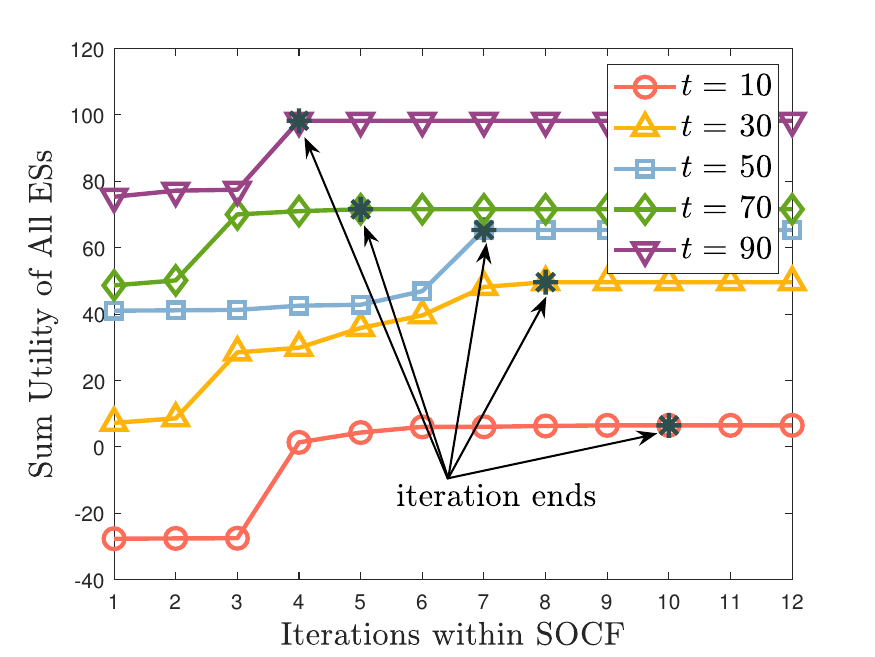}\\
	\vspace{-0.3em}
	\caption{Convergence of the designed SOCF algorithm.} \label{simu1}
	\vspace{-1.2em}
\end{figure}

Fig. \ref{simu0} illustrates the convergence of the proposed DMO approach. To assess the long-term performance, we define the cumulative system utility as $\sum\nolimits_{t = 0}^{T}\mathcal{U}(t) - C^{Conf}\mathcal{F}(t) -\max\big\{\psi \big (\tau^{Total}(t)-\delta_T \big ), 0\big\}$, where $T$ is set to 100 time frames. The results show that the system utility converges rapidly, indicating that all strategies can quickly converge to the equilibrium in the formulated hierarchical game $\mathcal{G}^H$. This is because of not only the use of virtual agents in the DRL, which reduce the complexity of both state and action spaces to enhance training efficiency, but also the adoption of PPO in the strategy exploration, which accelerates the convergence. All these findings align with the theoretical analysis presented in Theorem 1 and demonstrate the efficiency of the proposed DMO approach.

\textcolor{black}{Fig. 5 examines the convergence of the designed SOCF algorithm, showing that the total utility of all ESs, i.e., $\sum_{b \in \mathcal{B}} U_b(t)$ in each time frame $t$, can converge within a limited number of iterations, and the cumulative sum of all ESs' utilities $\sum_{t'=1}^t\sum_{b \in \mathcal{B}} U_b(t')$ increases as $t$ increases. This indicates that the equilibrium of the overlapping coalition formation subgame $\mathcal{G}^{Low}_t$ can be reached within each time frame. The reason is that no sensor has the motivation to deviate from its coalition once the final coalition partition, $\mathcal{CP}^*(t)$, is achieved when the iteration of SOCF algorithm terminates, which well matches the theoretical analyses of Lemma 2. Moreover, it is evident that the total number of iterations required to reach $\mathcal{CP}^*(t)$ decreases as $t$ increases. This is due to the diminishing marginal benefit of additional distributed sensing data in improving partial-DT model quality. As a result, when a substantial amount of historical data is available, ESs are less likely to associate with additional sensors (i.e., being reluctant to make frequent changes of the coalition partition), leading to a reduction in iterations of SOCF algorithm.}

\begin{figure}[!t]
	\centering
	\subfloat[]{
		\includegraphics[width=0.8\linewidth]{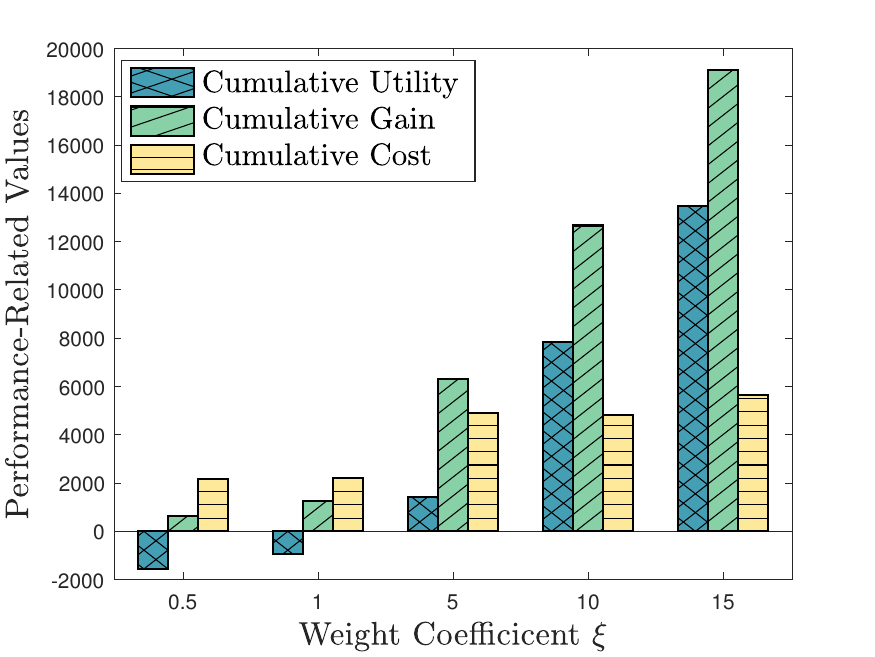}} 
	\vspace{-0.6mm}
	\quad  
	\subfloat[]{
		\includegraphics[width=0.8\linewidth]{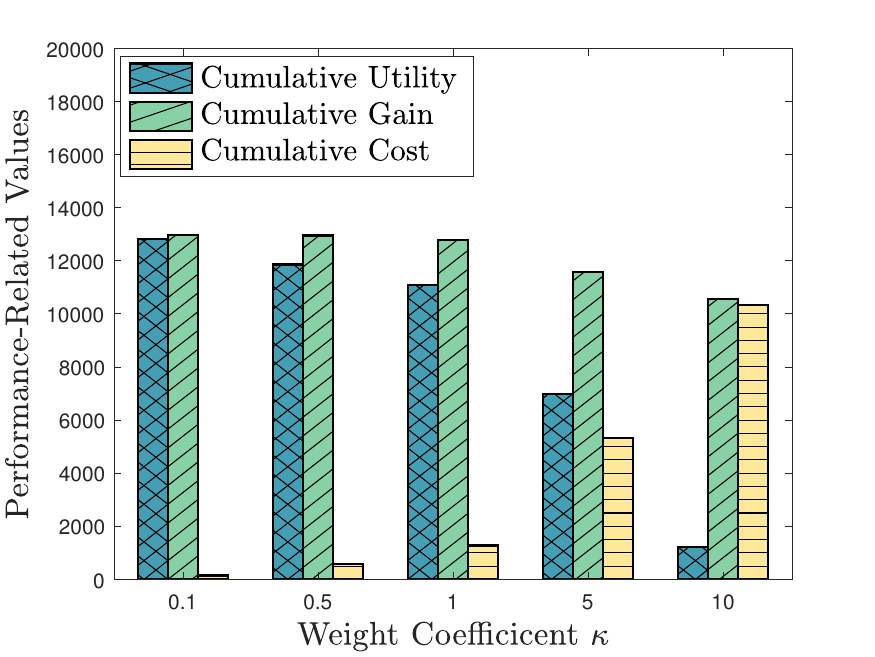}}
	\vspace{-0.3em}
	\caption{Long-term performance with different $\xi$ and $\kappa$.}
	\label{simu2}
	\vspace{-1.5em}
\end{figure}
\begin{figure}[!t]
	\centering
	\includegraphics[width=0.8\linewidth]{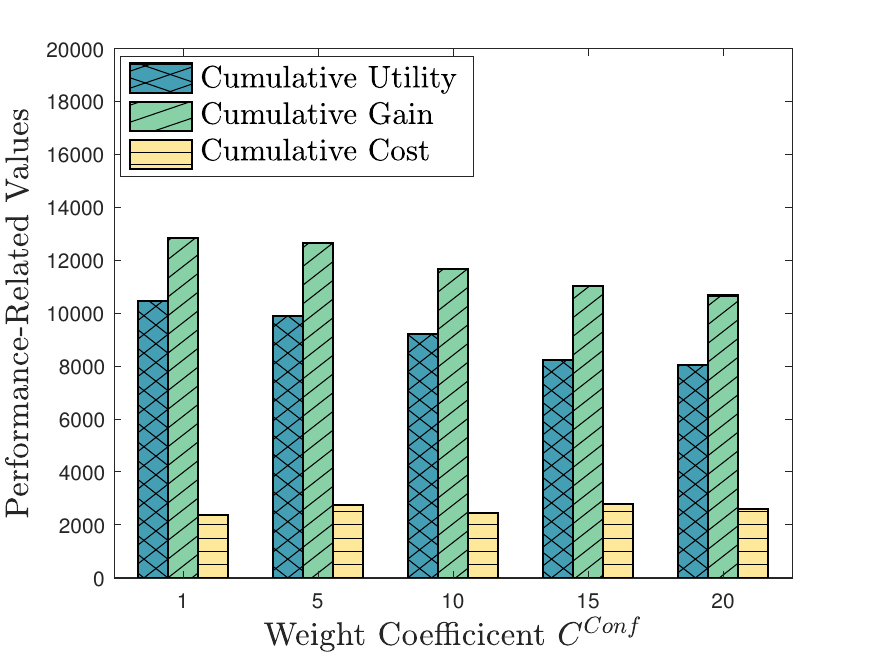}\\
	\vspace{-0.3em}
	\caption{Long-term performance with different $C^{Conf}$.} \label{simu3}
	\vspace{-1.2em}
\end{figure}
\begin{figure*}[tbp]
	\centering
	\subfloat[]{\label{simu4:a}\includegraphics[width=0.33\textwidth]{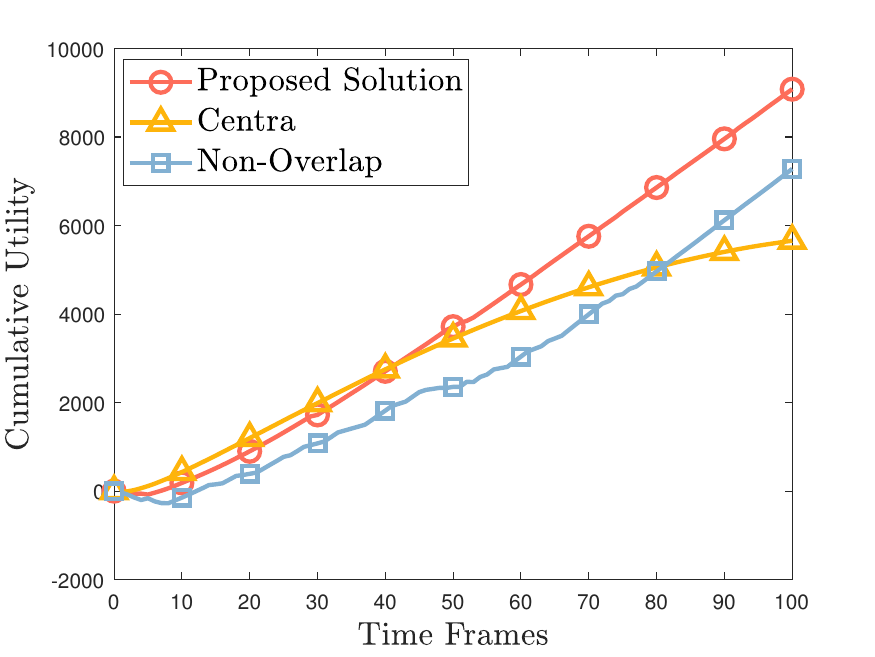}}
	\subfloat[]{\label{simu4:b}\includegraphics[width=0.33\textwidth]{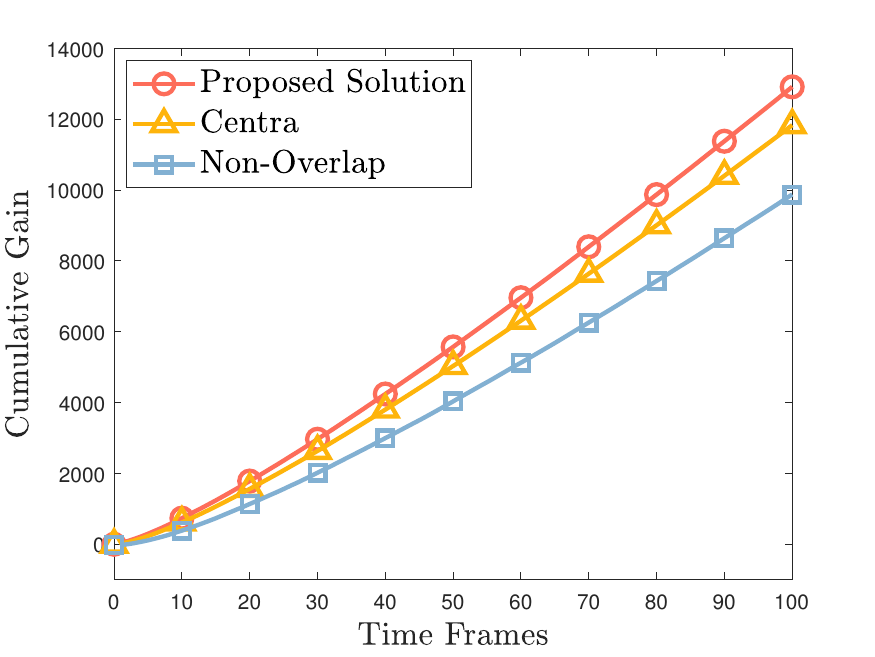}}	
	\subfloat[]{\label{simu4:c}\includegraphics[width=0.33\textwidth]{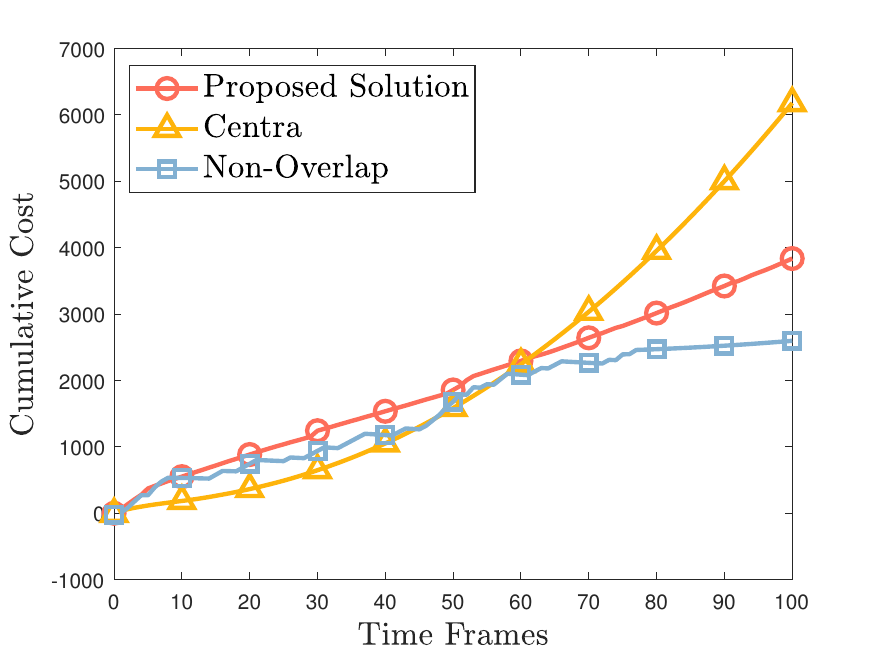}}
	\vspace{-0.5em}
	\caption{Performance comparison on different DT construction frameworks.}
	\label{simu4}
	\vspace{-1.9em}
\end{figure*}
\begin{figure*}[tbp]
	\centering
	\subfloat[]{\label{simu5:a}\includegraphics[width=0.33\textwidth]{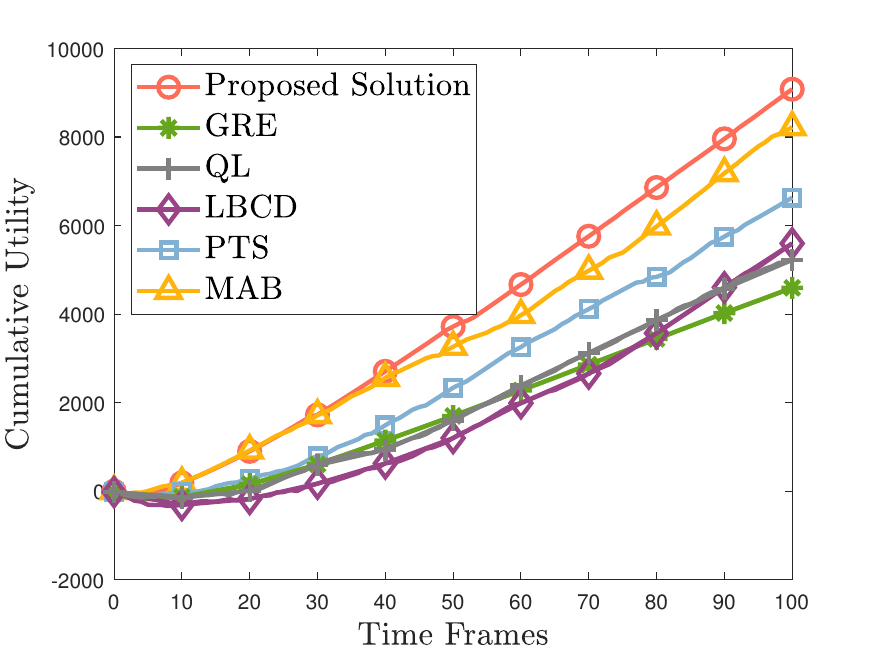}}
	\subfloat[]{\label{simu5:b}\includegraphics[width=0.33\textwidth]{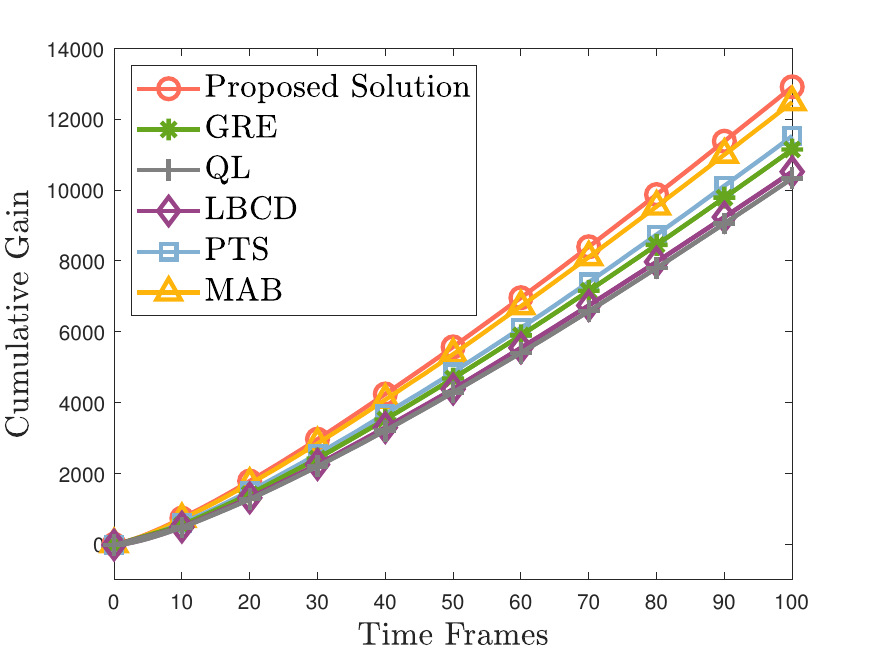}}	
	\subfloat[]{\label{simu5:c}\includegraphics[width=0.33\textwidth]{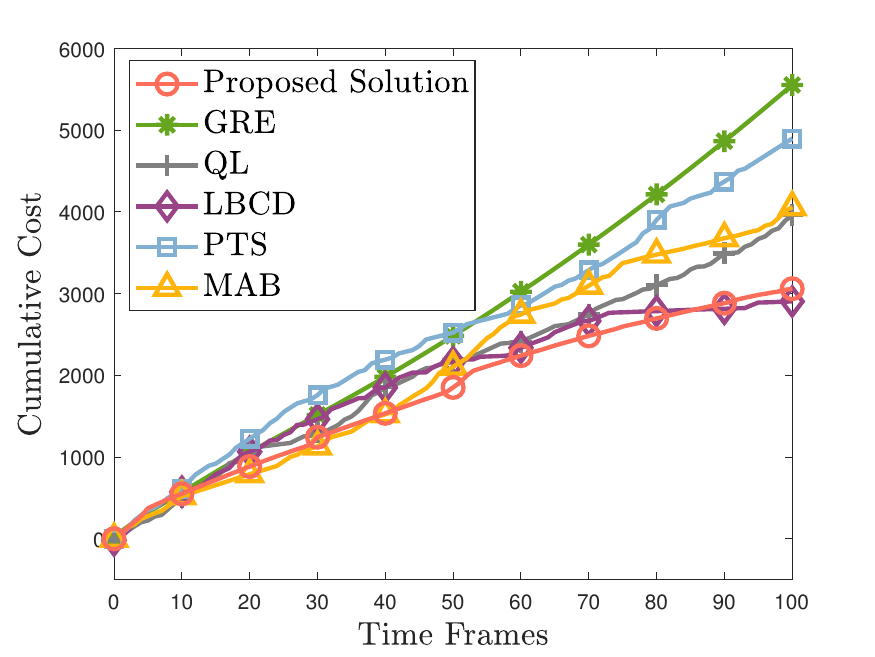}}
	\vspace{-0.5em}
	\caption{Performance comparison on different optimization methods.}
	\label{simu5}
	\vspace{-1.2em}
\end{figure*}
Fig. \ref{simu2} demonstrates the performance of federated DT construction using the proposed DMO approach, and show how long-term system performance $\mathcal{U}_{Sys}$, is influenced by weight coefficients $\xi$ and $\kappa$.
In Fig. \ref{simu2}(a), it is observed that as $\xi$ increases, both the cumulative utility and gain rise, while the cumulative cost tends to be stable. 
This occurs because, with a larger $\xi$, associating more sensors, i.e., collecting more feature data, significantly improves the quality of the global DT model. 
However, since the number of sensors that can be associated with each ES is limited, the cumulative cost is constrained.
In Fig. \ref{simu2}(b), while the cumulative cost sharply increases with $\kappa$, the DMO approach can still produce a relatively large system utility. This is because the DMO approach dynamically associates sensors with lower costs (e.g., communication overheads and energy consumptions) in data collection. Additionally, the cumulative gain in Fig. \ref{simu2}(b) is noticeably lower than that in Fig. \ref{simu2}(a) because, as $\kappa$ becomes sufficiently large, sensor associations are limited by the overwhelming cost, thereby restricting potential gains.


Fig. \ref{simu3} illustrates the impact of configuration cost $C^{Conf}$ on system's cumulative utility when the proposed DMO approach is employed.
It can be seen that, as $C^{Conf}$ increases, the cumulative gain gradually decreases.
This is because, when $C^{Conf}$ becomes sufficiently large, ESs are less willing to associate with additional sensors, i.e., dynamic sensor association $y_{b,n}(t)$ is constrained to a fixed one. This restriction limits the amount of feature data that each ES can acquire, and thus deter further improvements in model quality.
Moreover, we can observe that the cumulative cost does not increase drastically. 
This is because the designed SOCF algorithm can result in less frequent coalition changes for each ES if $C^{Conf}$ is considerably large.

Fig. \ref{simu4} demonstrates the effectiveness of the proposed federated DT construction framework over other DT construction schemes.
It can be observed from Fig. 8(a) and Fig. 8(b) that the proposed framework outperforms both $Centra$ and $Non\mbox{-}Overlap$ in terms of the cumulative utility and gain.
This is because the proposed framework allows sensors to be associated in an overlapped manner, enabling more efficient feature data collection, and thus results in a higher quality of the global DT model. Besides, although $Centra$ allows sensors to directly upload their data to the cloud, the number of sensors that can be associated is limited by the available subcarriers and strict delay constraints, leading to a lower cumulative gain.
Fig. 8(c) shows that, the proposed framework achieves lower cumulative costs than $Centra$ and only slightly higher costs than $Non\mbox{-}Overlap$. 
This is due to the fact that i) in $Centra$, the centralized data collection and DT creation introduce significant communication and computation latency, and ii) in $Non\mbox{-}Overlap$, while associating sensors without overlaps reduces the energy consumption for data transmission, compared with partial-DT creations and global DT integration that are relatively more resource-intensive, this actually accounts for only a small portion of the total energy consumption, which can hardly offset the loss of cumulative gain.

\textcolor{black}{Fig. 9 compares the proposed DMO approach with existing optimizing methods in terms of the long-term system performance. We can see that i) in Fig. 9(a), the proposed approach outperforms all benchmarks in terms of the cumulative utility, while $GRE$ exhibits the worst performance; ii) in Fig. 9(b), the cumulative gain of the proposed solution approach is significantly lower than that of all other benchmarks, while $QL$ exhibits the worst gain; and iii) in Fig. 9(c), the cumulative cost of the proposed solution approach is significantly lower than that of all other benchmarks except $LBCD$. These is because i) although $GRE$ can achieve a relatively high cumulative gain, it suffers from excessive costs due to greedily prioritizing the ES with the most feature data to create the partial-DT with the highest importance, ignoring the resulted energy consumptions and communication overheads; ii) $LBCD$ requires relaxation of originally discrete decision variables and an approximate decomposition of the long-term objective function, leading to a near-optimal solution when facing unpredictable system dynamics in federated DT construction; iii) although the proposed DMO approach and $QL$ are all based on reinforcement learning, they achieve different performance because the proposed DMO approach can better approximate the long-term non-convex cumulative utility through the neural network based actor and critic networks, rather than a predefined table in $QL$; iv) the proposed DMO approach have stronger exploration capabilities compared to heuristic algorithms (i.e., $PTS$) in interacting with the dynamic environment, reducing the possibility of falling into the local optima; v) unlike $MAB$, the proposed approach can fully leverage historical training data through the replay buffer of each agent, enhancing the long-term system performance.
}

\section{Conclusion}\label{CO}
In this paper, we have studied a federated DT construction framework enabled by \textcolor{black}{distributed} sensing under edge-cloud collaboration. We have formulated an online optimization problem to jointly and dynamically determine partial-DT assignments, ES-sensor associations, and as well as computation and communication resource allocations, for maximizing the long-term system performance. To tackle this complicated problem, we have conducted a transformation, forming an equivalent two-layer hierarchical game, consisting of an upper-layer two-sided matching game and a lower-layer overlapping coalition formation game.
After analyzing these games in detail, we have applied the GS algorithm and particularly developed an SOCF algorithm to respectively derive their short-term equilibria, and then proposed a DMO approach to accommodate dynamic settings, thereby solving the original problem.
Simulation results show the effectiveness of the federated DT construction framework and demonstrate the superiority of the proposed DMO approach over counterparts.

\textcolor{black}{ In the future work, we will further explore impacts brought by the non-i.i.d. data across devices in the federated DT construction. Particularly, the distributed sensing data may contain personalized information stems from unique characteristics of individual sensors, leading to significant partial-DT model deviations when determining ES-sensor associations in an overlapping manner. This may require a non-overlapping coalitional game for solving ES-sensor association, ensuring that each sensor is exclusively assigned to a single ES to maintain model consistency. Furthermore, the distributed sensing data may include distortions result from unpredictable sensing failures. To mitigate this, the partial-DT model replication and backup systems may be explored to improve the construction reliability.
}


\bibliographystyle{IEEEtran}
\bibliography{IEEEabrv,ref}
\newpage
\clearpage
\newpage
\appendices
\section{Proof of Lemma 1}
Since running the lower-layer subgame $\mathcal{G}^{Low}_t$ requires the partial-DT assignments from the upper-layer game $\mathcal{G}^{Up}_t$, a leader-follower relationship exists between these two games. In other words, the existence of the equilibrium in $\mathcal{G}^{Up}_t$ depends on that in $\mathcal{G}^{Low}_t$ \cite{OCFG1, BG2}.


Suppose the equilibrium of $\mathcal{G}^{Low}_t$ exists, $\mathcal{G}^{Up}_{t}$ is an one-to-one matching game with i) a finite set of participants $\mathcal{C}$ and $\mathcal{B}$, where the total amount is $|\mathcal{C}|+\mathcal{B}$; and ii) a finite set of strategy $\Pi^{U\hspace{-0.2mm}p}_t = \{\varPhi_t(i)\}_{\forall i\in G^{U\hspace{-0.2mm}p}}$, where the maximum cardinality is ${|\mathcal{B}|\choose |\mathcal{C}|} \cdot |\mathcal{C}|!$. Therefore, $\mathcal{G}^{Up}_{t}$ is a finite game which definitely has a Nash equilibrium according to \cite{47}.

\section{Proof of Lemma 2}
First, we prove that, given the optimal resource allocations $\{\pi^{R*}_b\}_{\forall b}$, the stable coalition partition $\mathcal{CP}^*(t)$ exists.
Considering that coalitions are formed in an overlapped manner, there are three possible cases for each sensor $n\in \mathcal{N}$ to be determined in changing its coalition, i.e., i) $n$ leaves one of its current coalitions and joins another one; ii) $n$ directly joins another coalition as an overlap; and iii) $n$ chooses to leave one of its current coalitions unilaterally. In each case, we prove that $\mathcal G^{Low}_{t}$ is an exact potential game (EPG) \cite{48}, which ensures the  existence of at least one pure Nash equilibrium or equivalently a stable overlapping coalition partition.
To this end, we can design a potential function as
\begin{align}
	\zeta^t(Co_a,Co_{-a})=\sum\nolimits_{b \in \mathcal{B}} U_b(t),\notag
\end{align}
where $Co_{-a}$ is set of all sensor coalitions of ESs other than $a\in \mathcal{B}$.
Hereafter, we analyze all three cases for each sensor $n$ and show that the change in the coalitional utility function $U_n(t)$ equals the change in potential function $\zeta^t$.

\emph{Case 1:} Suppose that sensor $n\in \mathcal{N}$ is determined to leave coalition $Co_a$ and join another coalition $Co_{a'}$, $\forall a, a'\in \mathcal{B}$, the difference brought to the utility of $n$, i.e., $U_{n}(t)$, can be calculated as
\begin{equation}
	\begin{aligned}
		&U_{n}(t) \big | Co_a, Co_{a'}-U_{n}(t) \big | Co_a\backslash n, Co_{a'}\cup n \notag \\
		=&U^{Co_a}_n(t) + U^{others}_n(t) -\big(U^{Co_{a'} \cup n}_n(t) + U^{others}_n(t)\big) \notag\\
		=&U^t_a(Co_a)-U^t_a(Co_a\backslash n) - \big(U^t_{a'}(Co_{a'}\cup n)-U^t_{a'}(Co_{a'})\big),\notag\\
		=&\sum\nolimits_{b\in \mathcal{B}} U_b(t) \big | Co_a, Co_{a'} - \sum\nolimits_{b\in \mathcal{B}} U_b(t) \big | Co_a\backslash n, Co_{a'}\cup n,\notag\\
		=&\zeta^t(Co_a,Co_{a'}, Co_{-a,a'}) - \zeta^t(Co_a\backslash n,Co_{a'}\cup n,Co_{-a,a'}),\notag
	\end{aligned}
\end{equation}
where $U^{others}_n(t)$ represents the total utility that $n$ obtains in other coalitions, $U^t_{-a,a'}$ and $Co_{-a,a'}$ denote the total utility and all coalitions of ESs other than $a$ and $a'$, respectively.

\emph{Case 2:} Suppose that sensor $n\in \mathcal{N}$ is determined to join coalition $Co_a,\forall a \in \mathcal{B}$ as an overlap, the difference brought to $U_{n}(t)$ can be calculated as
\begin{equation}
	\begin{aligned}
		&U_{n}(t)-U_{n}(t) \big | Co_a \cup n \notag \\
		=&U^{others}_n(t) -\big(U^{Co_{a} \cup n}_n(t) + U^{others}_n(t)\big) \notag\\
		=&U^t_a(Co_a)-U^t_a(Co_a \cup n),\notag\\
		=&U^t_a(Co_a)+U^t_{-a} -\big (U^t_a(Co_a \cup n) +U^t_{-a} \big ),\notag\\
		=&\sum\nolimits_{b\in \mathcal{B}} U_b(t) \big | Co_a - \sum\nolimits_{b\in \mathcal{B}} U_b(t) \big | Co_a \cup n,\notag\\
		=&\zeta^t(Co_a,Co_{-a}) - \zeta^t(Co_a\cup n,Co_{-a}).\notag
	\end{aligned}
\end{equation}

\emph{Case 3:} Suppose that sensor $n\in \mathcal{N}$ is determined to unilaterally quit its current coalition $Co_a,\forall a \in \mathcal{B}$. Obviously, this is equivalent to that sensor $n$ is determined to leave $Co_a$ and join an empty coalition $\varnothing$, and the difference in $U_n(t)$ equals to that in $\zeta^t$ following the proof in case 1.

Therefore, following \cite{48}, the lower-layer subgame $\mathcal{G}^{Low}_t$ is an EPG. Since it is widely proved that EPG has at least one pure Nash equilibrium, we can conclude that the equilibrium of $\mathcal{G}^{Low}_t$ (i.e., the stable overlapping coalition partition $\mathcal{CP}^*(t)$) in each time frame exists.


Finally, we prove that the best response $\big (\mathcal{CP}^*(t), \{\pi^{R*}_b\}_{\forall b}\big )$ exists. Given a stable coalition partition $\mathcal{CP}^*(t)$, the number of possible resource allocation strategies $\pi^{R*}_b$ for each ES $b \in \mathcal{B}$ is 
$|\mathcal{N}||\mathcal{W}| \hspace{-0.5mm} \left \lceil \hspace{-0.5mm} \frac{2 }{(L\delta-2)\delta \gamma} \log_2(1 \hspace{-0.5mm} - \hspace{-0.5mm} \frac{A_c^{Req}}{ \varGamma \big (\sum\nolimits_{i=1}^t \min_{n,c} d_{n,c}(t) \big)}) \hspace{-0.5mm} \right \rceil$,
implying that $\mathcal{G}^{Low}_t$ is also a finite game. 
According to \cite{47}, $\mathcal{G}^{Low}_t$ always has a Nash equilibrium, i.e., $\big (\mathcal{CP}^*(t), \{\pi^{R*}_b\}_{\forall b}\big )$. This completes the proof.
\section{Proof of Theorem 2}
In the proposed SOCF algorithm, each sensor $n \in \mathcal{N}$ is asked to perform the switch operations, i.e., $\mathcal{T}_n(Co_a, Co_b), \mathcal{J}_n(Co_b)$ or $\mathcal{Q}_n(Co_a)$, following the switch rules defined in Definition 7. Denote that the coalition partition becomes $\mathcal{CP}^{(k)}(t)$ after $k$ iterations, we prove that the sum utility of all ESs, i.e., $\sum_{\forall b \in \mathcal{B}}$, will not decrease from $\mathcal{CP}^{(k-1)}(t)$ to $\mathcal{CP}^{(k)}(t)$, regardless of which switch operation that sensor $n$ is determined to perform.

\emph{Case 1:} Suppose sensor $n\in \mathcal{N}$ is determined to perform the transferring operation $\mathcal{T}_n(Co_a, Co_b), \forall a, b\in \mathcal{B}$ in $k$-th iteration. Then, according to the transferring rule, we have
\begin{equation}
	\begin{aligned}
		&~U^{Co_b \cup n}_{n}(t) \geq \max\{0, U^{Co_a}_{n}(t)\} \\
		\Leftrightarrow&~U^t_b(Co_b \cup n) - U^t_b(Co_b) \geq  U^t_b(Co_a) - U^t_b(Co_a \backslash n) \\
		\Leftrightarrow&~U^t_b(Co_b \cup n) + U^t_b(Co_a \backslash n) \geq  U^t_b(Co_a) + U^t_b(Co_b).\notag
	\end{aligned}
\end{equation}
This implies that the total utility of coalitions $Co_a$ and $Co_b$ increases if $\mathcal{T}_n(Co_a, Co_b)$ is conducted.

\emph{Case 2:} 
Suppose that sensor $n\in \mathcal{N}$ is determined to perform the joining operation $\mathcal{J}_n(Co_b), \forall b \in \mathcal{B}$ in $k$-th iteration. Then, according to the joining rule, we have
\begin{equation}
	\begin{aligned}
		U^{Co_b \cup n}_{n}(t) \geq 0 
		\Leftrightarrow U^t_{b}(Co_{b}\cup n) \geq U^t_{b}(Co_{b}).\notag
	\end{aligned}
\end{equation}
This indicates that the utility of coalition $Co_b$ increases if $\mathcal{J}_n(Co_b)$ is conducted.

\emph{Case 3:} 
Suppose that sensor $n\in \mathcal{N}$ is determined to perform $\mathcal{Q}_n(Co_a), \forall a \in \mathcal{B}$ in $k$-th iteration. Then, according to the quitting rule, we have
\begin{equation}
	\begin{aligned}
		U^{Co_a}_{n}(t) \leq 0 
		\Leftrightarrow U^t_{a}(Co_{a}) \geq U^t_{a}(Co_{a}\backslash n). \notag
	\end{aligned}
\end{equation}
This shows that the utility of coalition $Co_a$ does not decrease if $\mathcal{Q}_n(Co_a)$ is conducted.

All these three cases demonstrate that whenever a sensor $n\in \mathcal{N}$ is determined to perform a coalition switch operation, the total utility of the coalitions either increases or maintains the same. Additionally, the utilities of all other coalitions will not decrease following Definition 7. Therefore, we can conclude that the total utility of all ESs, i.e., $\sum_{\forall b \in \mathcal{B}} U_b(t)$, monotonically increases in any iteration $k$ of SOCF algorithm.
Since the total utility of all ESs is bounded by the limited feature data that can be collected in each time frame, the coalition partition $\mathcal{CP}^{(k)}(t)$ is guaranteed to converge after a finite number of iterations.

\section{Proof of Theorem 3}
The computational complexity of the DMO approach primarily arises from the PPO-based training process integrated with the execution of GS and SOCF algorithms.

In each training step, the complexity of a forward pass through the actor or critic network of each agent with 3 fully connected layers is $\mathcal{O}(3K^2_{hid})$, where $K^2_{hid}$ denotes the number of hidden neurons in each layer. Meanwhile, each training step also involves a back propagation with the same computational complexity $\mathcal{O}(3K^2_{hid})$. 
Meanwhile, updating network parameters requires computing the PPO loss, which involves evaluating the value functions for all states and actions of two agents in MDP $\mathcal{M}_{PL}$ and $|\mathcal{B}|$ agents in MDP $\mathcal{M}_{R}$ across all time frames. Thus, the computational complexity is $\mathcal{O}\big (  2T(|\mathcal{S}_{PL}|+|\mathcal{A}_{PL}|)+|\mathcal{B}|T(|\mathcal{S}_{R}|+|\mathcal{A}_{R}|) \big )$.

It has been widely proved that the computational complexity of the well-known GS algorithm is $\mathcal{O}\big(|\mathcal{C}||\mathcal{B}|)\big)$. 
Besides, in SOCF algorithm, the number of coalitions that a sensor can simultaneously join is at most $L_n$, resulting in at most $L_n(|\mathcal{B}| - L_n)$ attempts of transfer operations. Therefore, the computational complexity of the SOCF algorithm is $\mathcal O \big(|\mathcal{N}|L_n(|\mathcal{B}|-L_n)\big)$. 

To sum up, the computation complexity of DMO can be expressed as
$\mathcal{O}\big(\textbf{I} \big( 2T(|\mathcal{S}_{PL}|+|\mathcal{A}_{PL}|)+|\mathcal{B}|T(|\mathcal{S}_{R}|+|\mathcal{A}_{R}|) + 6K^2_{hid}  + |\mathcal{C}||\mathcal{B}| + \varpi |\mathcal{N}|L_n(|\mathcal{B}|-L_n) \big) \big)$.

\vfill

\end{document}